\documentclass[5p, twocolumn]{elsarticle}

\usepackage{lineno,hyperref}

\journal{Journal of Power Sources}

\usepackage{todonotes}
\usepackage{amsmath}
\usepackage{placeins}
\usepackage{booktabs}
\usepackage{lipsum}
\usepackage{supertabular}

\usepackage{color}

\begin{document}

\begin{frontmatter}

\title{On-board monitoring of 2-D spatially-resolved temperatures in cylindrical lithium-ion batteries: Part I. Low-order thermal modelling}

\author{Robert R. Richardson, Shi Zhao and David A. Howey\fnref{myfootnote}}
\address{Department of Engineering Science, University of Oxford, Oxford, UK}
\fntext[myfootnote]{E-mail: \{robert.richardson, shi.zhao, david.howey\} @eng.ox.ac.uk.}

\begin{abstract}
	Estimating the temperature distribution within Li-ion batteries during operation is critical for safety and control
	 purposes. Although existing control-oriented thermal models - such as thermal equivalent circuits (TEC) - are computationally efficient, they only predict average temperatures, and are unable to predict the spatially resolved temperature distribution throughout the cell.
	We present a low-order 2D thermal model of a cylindrical battery based on a Chebyshev spectral-Galerkin (SG) method, capable of predicting the full temperature distribution with a similar efficiency to a TEC.
	The model accounts for transient heat generation, anisotropic heat conduction, and non-homogeneous convection boundary conditions.
	The accuracy of the model is validated through comparison with finite element simulations, which show that the 2-D temperature field ($r, z$) of a large format ($64$ mm diameter) cell can be accurately modelled with as few as 4 states. 
	Furthermore, the performance of the model for a range of Biot numbers is investigated via frequency analysis. For larger cells or highly transient thermal dynamics, the model order can be increased for improved accuracy.
	The incorporation of this model in a state estimation scheme with experimental validation against thermocouple measurements is presented in the companion contribution (Part II).
\end{abstract}

\begin{keyword}
Lithium-ion battery\sep thermal model\sep spectral methods\sep low-order modelling\sep temperature estimation
\end{keyword}

\end{frontmatter}


\section*{Highlights}
\begin{itemize}
	\item{Derivation and validation of low-order 2-D thermal model for cylindrical cells.}
	\item{Efficient numerical implementation based on Chebshev spectral-Galerkin method.}
	\item{Includes anisotropic heat conduction and inhomogeneous convection boundary conditions.}
	\item{Applicable to various battery cooling configurations, such as side or end cooling.}
	\item{Suitable for use in a state-estimator with surface temperature or EIS measurements.}
\end{itemize}

\section{Introduction\label{sec:Introduction}}

Lithium-ion batteries generate heat due to electrochemical processes, which results in internal temperature gradients during operation.
In a typical usage scenario, such as a standard vehicle drive cycle, cells may experience temperature differences between surface and core of 20  $^{\circ}$C or more \cite{Forgez2010a}; and during a rapid overheating event this discrepancy can be as large as 40-50 $^{\circ}$C \cite{Spinner2015}. High battery temperatures could trigger thermal runaway resulting in fires, venting and electrolyte leakage. While such incidents are rare \cite{Wang2012a}, consequences include costly recalls and potential endangerment of human life. Consequently, transient thermal modelling of batteries during operation is an essential requirement for battery management systems to ensure safe and optimal performance.

In this study, we present and validate a low-order thermal model of a cylindrical battery cell, capable of capturing 2-D thermal dynamics. The model is based on the spectral-Galerkin method, achieving high accuracy with minimal computational requirements, making it suitable for online applications. The remainder of this paper is organised as follows. In Section 2, the alternative numerical methods for low-order thermal modelling are discussed. In Section 3, a gentle introduction to Galerkin-spectral methods is provided by means of a toy problem - a 1-D heat equation in Cartesian coordinates. In Section 4, the full 2-D thermal model is presented; and in Section 5, the results of the model are validated through comparison with high fidelity Finite Element (FE) simulations. {Matlab code to simulate the presented model is available online\footnote{{www.github.com/robert-richardson/Spectral-Thermal-Model-2D}}.}

\section{Low-order thermal modelling}

Lumped parameter thermal equivalent circuit (TEC) models are perhaps the most popular approach for efficient thermal modelling. These methods have been used extensively for low-order and control-oriented modelling of battery cells and packs \cite{Forgez2010a, Li2013a, Lin2011, Lin2014, Mahamud2011a, Damay2013, Dai2015, He2015}. Their main advantage is that they are simple to implement. However, they are unable to predict the full temperature field throughout the domain of interest; their outputs merely consist of nodal values representing average temperatures. This is a particular limitation when comparison with temperature measurements at discrete locations is necessary for state or parameter estimation. Moreover, since the parameters of the model have no physical meaning, they require parameterization using experimental data for any given set of parameters and operating conditions. On the other hand, TEC models with parameters that can be directly calculated from physical properties have been used extensively in thermal modelling of electric machines and other applications \cite{mellor1991lumped,Wrobel2010,simpson2014general}. However, these are known to have poor performance for large Biot numbers, whilst increasing the number of elements to improve accuracy comes at the expense of increased computational complexity \cite{qi2014methodical,simpson2014accurate}.

Physics based models instead solve the underlying diffusion partial differential equation (PDE) governing the heat transfer. They are generally applicable to a broad range of problems, and can predict the full temperature field throughout the domain of interest. Several studies have presented 2-D or 3-D thermal simulations of battery cells \cite{Kim2007, Pesaran2002d, Gu2000a, Srinivasan2003}]. However, the solution is typically obtained using computationally intensive numerical methods such as Finite Difference Methods (FDM) or Finite Element Methods (FEM), and so their potential for application in control systems is limited. Analytical solutions have also been developed \cite{Shah2014a, Shah2015modeling}, however these are inappropriate for on-line applications since time domain solutions rely on computationally intensive integral transforms.

\begin{figure}[h]
	\centering
	\includegraphics[width=0.95\columnwidth]{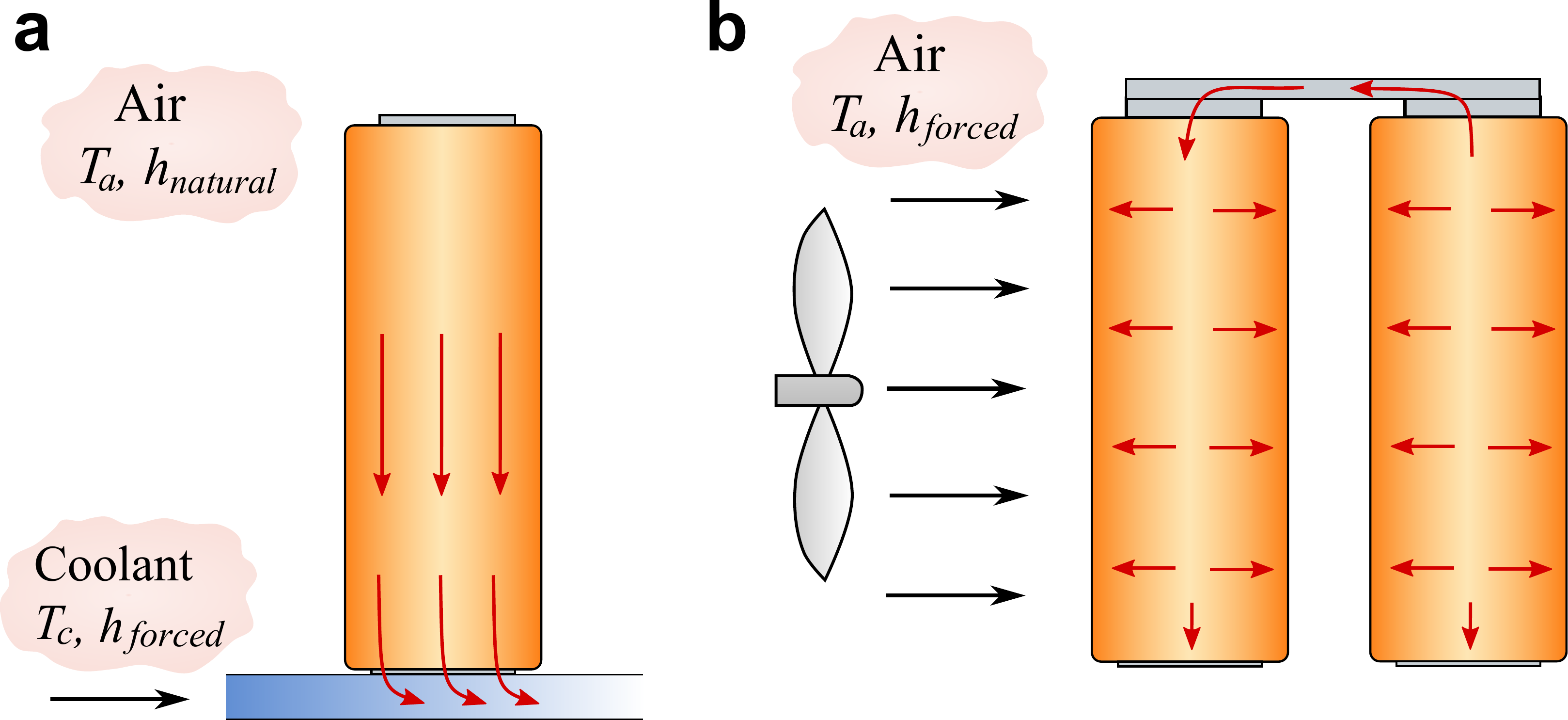}
	\caption{Battery cooling configurations resulting in 2-D thermal dynamics: (a) Plate cooling, (b) unequal cooling with intercell heat transfer}
	\label{fig:thermal2D}
\end{figure}

To reduce the computational burden of physics based models, low order approaches have been proposed  using techniques such as balanced truncation \cite{Muratori2010a, Muratori2010b} and polynomial approximation (PA) \cite{Kim2013, Kim2014b, Richardson2014, Richardson2015a}. However, these methods are only suited to 1-D problems involving infinite or semi-infinite domains, or symmetric boundary conditions~\cite{hahn2012heat}. This may be acceptable for cases in which thermal gradients arise predominantly in one direction, such as in air cooling of small form factor (e.g. 18650 or 26650) cylindrical cells. However, large form factor cells have a greater propensity for thermal gradients in multiple directions. Recent studies have used 45 Ah cylindrical cells with a diameter of 64 mm and height of 198 mm \cite{Roscher2015}. Such dimensions give rise to much larger Biot numbers and hence more significant thermal gradients. Moreover, certain cooling configurations, such as end cooling via cooling plates, are more likely to give rise to axial variations. These systems also introduce additional complexities, such as the potential for each surface of the cell to be in contact with a different cooling fluid with a different free stream temperature and/or convection coefficient. Fig. \ref{fig:thermal2D}a shows a case with forced convection via a liquid coolant at the base of the cell and natural convection to the air at the other surfaces. Fig. \ref{fig:thermal2D}b shows another scenario in which unequal cooling of multiple cells can result in 2-D thermal dynamics as heat is transferred axially through the tabs from one cell to a neighbouring cell. Consequently, there is a clear need for low-order thermal models capable of capturing 2-D thermal dynamics.

Spectral methods are an alternative numerical method for solving PDEs, in which the spatial discretization is carried out using global rather than local approximating functions \cite{Shen2011, trefethen2000spectral}. In general, FDM and FEM methods are suitable for complex geometries, whereas spectral methods provide greater computational efficiency at the expense of model flexibility and the assumption of a smooth solution. Spectral methods are a type of weighted residual method - a group of approximation techniques in which the solution errors are minimized in a certain way - and they are classified according to the minimization technique employed. The most common techniques are spectral-collocation and spectral-Galerkin. In a collocation method, the solution is obtained by interpolating an approximating function at a set of domain nodes, whereas in a Galerkin method, the solution is obtained by forcing the residual of an integral multiplied by a test function to zero.

Spectral methods have been used in previous studies for low-order battery modelling \cite{Subramanian2005,Northrop2011a, Cai2012a, bizeray2015lithium}, and recently they have even been applied to 2-dimensional problems \cite{Northrop2015}. However, to the authors' knowledge, no study has applied a Galerkin method to 2-D battery thermal problems. This is perhaps due to the complexity of accounting for non-homogeneous boundary conditions, in particular convection boundary conditions, using a Galerkin method.
One of the advantages of the Galerkin method is that the basis functions implicitly satisfy the boundary conditions, and so it possible to have very low order models with satisfactory accuracy.

In this paper, we present a 2-D model of a cylindrical cell based on a Chebyshev spectral-Galerkin (SG) method. The model accounts for transient heat generation, anisotropic heat conduction, and non-homogeneous convection boundary conditions, with different convection coefficients and external temperatures at each surface. This generality makes it suitable for simulating various battery cooling configurations, such as those discussed previously. The treatement of the non-homogeneous boundary conditions is achieved by an efficient boundary lifting algorithm, adapted from a recently developed boundary lifting algorithm for elliptic problems \cite{Doha2012a}. Since the underlying basis functions implicitly satisfy the boundary conditions, there is no need for additional equations to impose the boundary conditions. As a result accurate models can be generated using as a few as two basis functions in each direction, and thus $2^2 \equiv 4$ states. The accuracy can be improved arbitrarily by increasing the number of basis functions in either dimension.

\section{Toy problem\label{sec:SG_toy}}

Firstly, we introduce SG methods by means of a simple example - a 1-D heat equation in Cartesian coordinates on the domain $\hat{x} \in [-1, 1]$, with non-homogeneous boundary conditions.

\subsubsection*{Governing equation}

The governing equation is given by:
\begin{equation}
\rho c_{p}\frac{\partial T}{\partial t} - k\frac{\partial^{2}T}{\partial \hat{x}^{2}} - q = 0,
\label{eq:toy_1}
\end{equation}
where $t$ is time, $\hat{x} \in [-1, 1]$ is the position coordinate and $k$ is the thermal conductivity.
The convection boundary conditions are given by
\begin{align}
& a_+ T + b_+ \frac{\partial}{\partial \hat{x}} T  = e_{r}
& & \text{at } \hat{x} = 1
\\
& a_- T + b_- \frac{\partial}{\partial \hat{x}} T  = e_{l}
& & \text{at } \hat{x} = -1
\label{eq:toy_2}
\end{align}
where
\begin{align}
a_+ & = h/k \text{,} & & b_+  = 1 , \\
a_- & = -h/k \text{,} &  & b_-  = 1 , \\
e_r & = a_{+} T_{\infty} \text{,} &  & e_l  = a_{-} T_{\infty},
\end{align}
where $h$ is the convection coefficient and $T_{\infty}$ is the external temperature. For simplicity the convection coefficient and external temperature at each boundary are assumed to be the same.

\subsubsection*{Finite sum approximation}
The starting point of the SG method is to approximate the solution $T$ of eq.~(\ref{eq:toy_1}) by a finite sum
\begin{equation}
T = \sum\limits_{n=0}^{N} a_n \phi_n(\hat{x}) + T_{\infty},
\label{eq:toy_3}
\end{equation}
where $a_{n}$ are unknown solution coefficients, and $\phi_n$ are the \emph{trial} (\emph{or basis}) \emph{functions} which must satisfy the Robin boundary conditions of eq. (\ref{eq:toy_2}).

\subsubsection*{Basis functions}

{In principle any Jacobi polynomial may be used for the basis functions, however, for this example, Chebyshev functions were used since they can be conveniently defined such that they adhere to any of a broad class of boundary conditions~\cite{Shen2011}}. Suitable basis functions are found as follows. 
We denote by $C_n$ the $n$th degree Chebyshev polynomial of the first kind and then let $\{\phi_n(\hat{x})\}_{n=0}^N $ be a basis {function} such that 
\begin{equation}
\phi_n = C_n(\hat{x}) + \zeta_n C_{n+1}(\hat{x}) + \eta_n C_{n+2}(\hat{x}) \text{,}
\label{eq:toy_basis_1}
\end{equation}
where $\zeta_n$ and $\eta_n$ are defined according to the formula in Lemma 4.3 of \cite{Shen2011} such that they adhere to the boundary conditions,
\begin{equation}
\zeta_n = \{ 4 \left(n+1 \right) \left(a_+ b_- + a_- b_+ \right) \} / \mathrm{DET}_n,
\end{equation}
and
\begin{equation}
\begin{split}
\eta_n = 
\{ -2a_-a_+ + \left(n^2 + (n+1)^2\right) (a_+ b_- - a_-b_+)  \\
+ 2b_-b_+n^2(n+1)^2  \} / \mathrm{DET}_n,
\end{split}
\end{equation}
where
\begin{equation}
\begin{split}
\mathrm{DET}_n = 2 a_+ a_- + \left( (n+1)^2 +(n+2)^2\right) \\
(a_-b_+ - a_+b_-)
-2b_-b_+(n+1)^2(n+2)^2.
\end{split}
\label{eq:toy_basis_2}
\end{equation}

From this point on, $\phi_n$ are considered known functions.

\subsubsection*{Residual equation}

Substituting eq.~(\ref{eq:toy_3}) for $T$ into eq.~(\ref{eq:toy_1}) leads to  the \emph{residual}:
\begin{equation}
R:= \rho c_{p}\frac{\partial T}{\partial t} - k\frac{\partial^{2}T}{\partial \hat{x}^{2}} - q \neq 0.
\label{eq:toy_4}
\end{equation}
The principle of the SG method is to force an integral of the residual to zero by requiring
\begin{equation}
\int\limits_{-1}^{1} \left( \rho c_{p}\frac{\partial T}{\partial t} - k\frac{\partial^{2}T}{\partial \hat{x}^{2}} - q \right) \nu \mathrm{d}\hat{x}  = 0,
\label{eq:toy_6}
\end{equation}
where $\nu$ is a \emph{test function}.

Substituting eq.~(\ref{eq:toy_3}) for $T$ and $\phi_n$ for $\nu$ into eq.~(\ref{eq:toy_6}), we have
\begin{multline}
\rho c_p \begin{bmatrix}
\int_{-1}^{1}\phi_0 \phi_n\mathrm{d}\hat{x},\, \dots,\, \int_{-1}^{1}\phi_N \phi_n \mathrm{d}\hat{x}
\end{bmatrix}
\begin{bmatrix}
\frac{\mathrm{d}}{\mathrm{d}t} a_0\\
\vdots\\
\frac{\mathrm{d}}{\mathrm{d}t} a_N
\end{bmatrix}
\\
- k \begin{bmatrix} \int_{-1}^{1} \frac{\partial \phi_0}{\partial \hat{x}} \phi_n \mathrm{d}\hat{x}
,\, \dots,\, 
\int_{-1}^{1} \frac{\partial \phi_N}{\partial \hat{x}} \phi_n \mathrm{d}\hat{x}
\end{bmatrix}
\begin{bmatrix}
{a}_0\\
\vdots\\
{a}_N
\end{bmatrix}
\\
- q \int\limits_{-1}^{1} \phi_n \mathrm{d}\hat{x} = 0.
\end{multline}

\subsubsection*{State space equation}

With $n = 0,\,1, \dots ,\, N$, the above $N+1$ equations representing the spatial discretisation can be written in compact form as
\begin{equation}
\mathbf{E}\mathbf{\dot{x}} = \mathbf{A}\mathbf{x} + \mathbf{B}\mathbf{u},
\end{equation}
where
\begin{align}
\mathbf{x} & = \left[	a_0, \, \dots,\, a_N \right]',
\\
\mathbf{u} & = \left[	q, \, T_{\infty} \right]',
\end{align}
and
\begin{align}
\mathbf{E}(i,j) & = \rho c_p \int_{-1}^{1}\phi_i \phi_j \mathrm{d}\hat{x}, \\
\mathbf{A}(i,j) & = k \int_{-1}^{1} \frac{\partial^2 \phi_j}{\partial \hat{x}^2} \phi_i \mathrm{d}\hat{x}, \\
\mathbf{B}(i,1) & = \int_{-1}^{1} \phi_i \mathrm{d}\hat{x},\\
\mathbf{B}(i,2) & = 0.
\end{align}

Letting the outputs of the system be the temperature at the left and right boundaries, $T(\hat{x}=-1)$ and $T(\hat{x}=1)$, we have
\begin{equation}
\mathbf{y} = \mathbf{C}\mathbf{x} + \mathbf{D}\mathbf{u},
\end{equation}
where
\begin{align}
\mathbf{y} & = \left[ T_{\hat{x}=-1}, T_{\hat{x}=1} \right]^T, \\
C_{1,j} & = \begin{bmatrix}
\phi_0 (\hat{x} = -1) , \, \dots , \, \phi_N(\hat{x} = -1)
\end{bmatrix} ,\\
C_{2,j} & = \begin{bmatrix}
\phi_0 (\hat{x} = 1) , \, \dots , \, \phi_N(\hat{x} = 1)
\end{bmatrix},
\\
\mathbf{D} & = 
\begin{bmatrix}
0 & 1 \\
0 & 1
\end{bmatrix}.
\end{align}

As we shall see, the actual 2-D problem of interest presented in the following section is more complex than the above problem, for the following reasons: (i) it involves cylindrical rather than Cartesian coordinates, (ii) the physical domain is not necessarily in the range $[-1,1]$ and must be scaled accordingly, and (iii) it is 2-D, which renders the homogenisation step of the non-homogeneous boundary conditions non-trivial. However, the solution processes in both cases are equivalent.

\section{Thermal model\label{sec:SG_Model_Main}}

\subsection{Overview\label{sec:SG_Overview}}

We now describe the full 2-D SG model.
A step-by-step procedure is given as follows. In Section~\ref{sec:SG_Model}, the 2-D thermal problem is defined. In Section~\ref{sec:SG_Scale}, the model is scaled from the physical coordinates to a coordinate system suitable for implementation of the SG algorithm. In Section~\ref{sec:SG_Homog}, the scaled model is decomposed into a homogeneous problem and a (time-invariant) boundary lifting function. Section~\ref{sec:SG_Galerkin} presents the solution to the homogeneous problem using the SG method and Section~\ref{sec:SG_Boundary_lifting} presents the calculation of the boundary lifting function. Finally, Section~\ref{sec:SG_Sol} presents the overall solution to the original problem which combines the solutions obtained from Sections~\ref{sec:SG_Galerkin} and \ref{sec:SG_Boundary_lifting}. The overall procedure ultimately generates a state-space model; the resulting model can then be solved efficiently online.

\subsection{Model definition\label{sec:SG_Model}}

\begin{figure}[hbt]
	\centering
	\includegraphics[width=0.8\columnwidth]{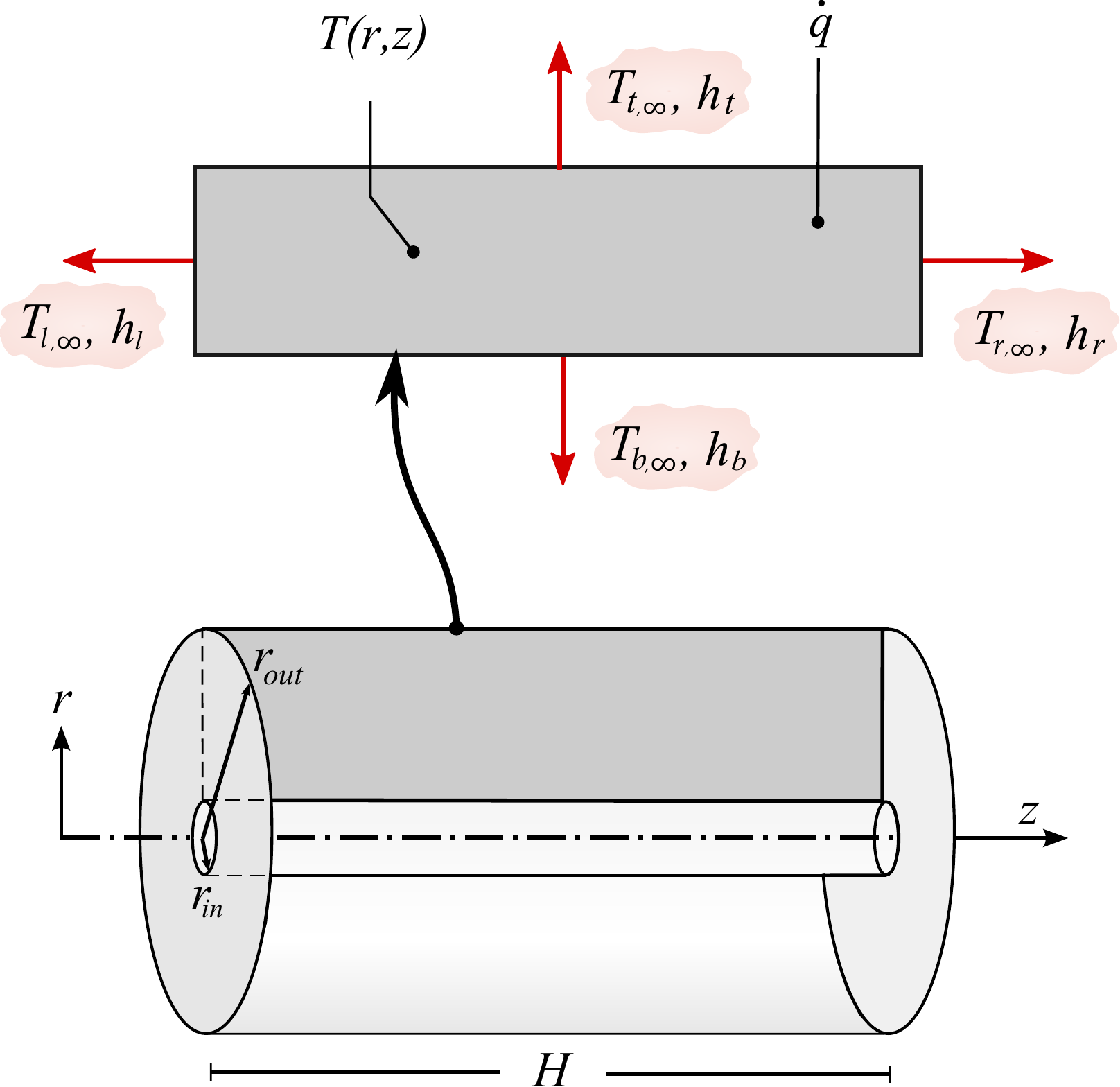}
	\par
	\caption[Schematic of cylindrical cell geometry for the thermal model]{Schematic of cylindrical cell geometry for the thermal model, showing different convection coefficients and/or external temperatures at each surface.}
	\label{fig:schematic}
\end{figure}

The model consists of the transient energy conservation equation in cylindrical coordinates. Heat generation is assumed to be uniform in space but time-dependant, but as we show in part II of this paper, the error introduced by this assumption is modest. The multi-layer structure of the battery is treated as a homogeneous solid with anisotropic thermal conductivity in the radial and axial directions. The temperature variation in the azimuthal ($\varphi$) direction is neglected. Convective heat transfer is assumed to occur at the outside surfaces, and the properties of the heat transfer fluid (i.e. the heat transfer coefficient and the fluid free-stream temperature) may be different for each surface (see Figure \ref{fig:schematic}).

The resulting model is governed by the following 2-D boundary value problem \cite{hahn2012heat}:
\begin{equation}
\rho c_{p}\frac{\partial T}{\partial t} - k_{r}\frac{\partial^{2}T}{\partial r^{2}} -
\frac{k_{r}}{r}\frac{\partial T}{\partial r} - k_{z}\frac{\partial^{2}T}{\partial z^{2}} = q
\label{eq:original_eq}
\end{equation}
where $t$ is time and $r$ and $z$ are the position coordinates in the radial and axial directions respectively. The functions $T(r,z,t)$ and $q(t)$ are the temperature distribution and volumetric heat generation rate, respectively. The parameters $\rho$ and $c_p$ are the density and specific heat capacity respectively, and $k_r$ and $k_z$ are the anisotropic thermal conductivities in the $r$ and $z$ directions. 
The boundary conditions are given by:
\begin{subequations}
	\begin{align}	
	\frac{\partial T}{\partial r} & =  -\frac{h_{r}}{k_{r}}(T-T_{t, \infty})
	\text{\quad at $r = r_{out}$} \label{eq:BC1_chap7} \\
	\frac{\partial T}{\partial r} & =  \frac{h_{l}}{k_{r}}(T-T_{b, \infty})
	\text{\quad at $r = r_{in}$} \\
	\frac{\partial T}{\partial z} & = 	-\frac{h_{t}}{k_{z}}(T-T_{r, \infty})
	\text{\quad at $z = H$} \label{eq:BC3}\\
	\frac{\partial T}{\partial z} & = 	\frac{h_{b}}{k_{z}}(T-T_{l, \infty})
	\text{\quad at $z = 0$} \label{eq:BC4}
	\end{align}
\end{subequations}
where $\{T_{\sigma, \infty}; \, \sigma = t, b, r \text{ and } l \}$ are the free-stream temperatures of the heat transfer fluid at the top, bottom, right and left surfaces\footnote{Note that although we have prescribed constant (w.r.t the spatial variable) external temperatures at each side, the model is also capable of handling spatially dependant temperatures (e.g. $T_{r,\infty}(z)$ or $T_{t,\infty}(r)$). However, since the external temperatures would rarely be known with such a high fidelity in a real application, we chose constant values for the sake of simplicity.}, and $\{h_{\sigma}; \, \sigma = t, b, r \text{ and } l \}$ are the corresponding convection coefficients.

\subsection{Change of scale\label{sec:SG_Scale}}
In order to exploit the properties of the polynomial basis functions it is necessary to scale from the original (physical) domain, $(r,\,z)$ where $r \in [r_{in}, r_{out}]$, $z \in [0, H]$, to the spectral domain, $(\hat{r},\hat{z})$ where $\hat{r} \in [-1, 1], \, \hat{z}\in [-1, 1]$. We define the spectral coordinates by
\begin{subequations}
	\begin{align}
	\hat{r} & = 2 \frac{r-r_{in}}{r_{out}-r_{in}} - 1
	\text{\quad s.t. }
	\begin{cases}
	\hat{r}(r = r_{in}) & = -1\\
	\hat{r}(r = r_{out}) & = 1
	\end{cases}
	\label{eq:scale_r1}\\
	\hat{z}  & = \frac{2 z}{H} - 1
	\text{\quad s.t. }
	\begin{cases}
	\hat{z}(z = 0) & = -1\\
	\hat{z}(z = H) & = 1
	\end{cases}
	\label{eq:scale_z1}
	\text{.}
	\end{align}
\end{subequations}
Rearranging (\ref{eq:scale_r1}) and (\ref{eq:scale_z1}), the original coordinates are given by
\begin{subequations}
	\begin{align}
	r & = \frac{1+\hat{r}+\alpha r_{in}}{\alpha},\label{eq:scale_r2}\\
	z & =  \frac{\hat{z} + 1}{\beta}.\label{eq:scale_z2}
	\end{align}
\end{subequations}
where $\alpha$ and $\beta$ are scaling factors defined by $\alpha = 2/ (r_{out}-r_{in})$ and $\beta = 2/H$.

Substituting (\ref{eq:scale_r2}) and (\ref{eq:scale_z2}) into the original eq. (\ref{eq:original_eq}), we obtain the governing equation in the spectral domain:
\begin{equation}
\rho c_{p}\frac{\partial T}{\partial t}
- \alpha^2 k_{r}\frac{\partial^{2}T}{\partial \hat{r}^{2}}
- \frac{\alpha^2 k_r}{1+ \alpha r_{in} + \hat{r}}\frac{\partial T}{\partial \hat{r}}
- \beta^2 k_{z}\frac{\partial^{2}T}{\partial \hat{z}^{2}}
= q
\label{eq:spectral_eq}
\end{equation}
with the boundary conditions:
\begin{subequations}
	\begin{align}
	& a_+ T + b_+ \frac{\partial}{\partial \hat{r}} T  = e_{t}
	& & \text{at } \hat{r} = 1
	\label{eq:BC1b}\\
	& a_- T + b_- \frac{\partial}{\partial \hat{r}} T  = e_{b}
	& & \text{at } \hat{r} = -1
	\label{eq:BC2b}\\
	& c_+ T + d_+ \frac{\partial}{\partial \hat{z}} T  = e_{r}
	& & \text{at } \hat{z} = 1
	\label{eq:BC3b}\\
	& c_- T + d_- \frac{\partial}{\partial \hat{z}} T  = e_{l}
	& & \text{at } \hat{z} = -1
	\label{eq:BC4b}
	\text{,}
	\end{align}
\end{subequations}
where,
\begin{align}
\nonumber & a_+ = h_r/k_r \text{,} & & b_+  = \alpha \text{,} \\
\nonumber & a_- = -h_l/k_r \text{,} & & b_-  = \alpha \text{,} \\
\nonumber & c_+ = h_t/k_z \text{,} & & d_+  = \beta \text{,} \\
\nonumber & c_- = -h_b/k_z \text{,} & & d_-  = \beta \text{,} \\
\nonumber & e_{t} = a_+ T_{t, \infty} \text{,} & & e_{b}  = a_- T_{b, \infty} \text{,} \\
& e_{r} = c_+ T_{r, \infty} \text{,} & & e_{l}  = c_- T_{l, \infty}.
\label{eq:BCconstants}
\end{align}

\subsection{Homogenization of the boundary conditions\label{sec:SG_Homog}}

Next, the problem with non-homogeneous boundary conditions is transformed into a problem with homogeneous boundary conditions using a boundary lifting algorithm, as follows:
We set
\begin{equation}
T = \tilde{T} + T_e \text{,}
\label{eq:boundary_auxiliary}
\end{equation}
where $T_e(\hat{r},\hat{z})$ is an arbitrary function satisfying the original boundary conditions (see later), and $\tilde{T}(\hat{r},\hat{z},t)$ is an auxiliary function satisfying the modified problem
\begin{equation}
\rho c_{p}\frac{\partial \tilde{T}}{\partial t}
- \alpha^2 k_{r}\frac{\partial^{2}\tilde{T}}{\partial \hat{r}^{2}}
- \frac{\alpha^2 k_r}{1+\alpha r_{in} + \hat{r}} \frac{\partial \tilde{T}}{\partial \hat{r}}
- \beta^2 k_{z}\frac{\partial^{2}\tilde{T}}{\partial \hat{z}^{2}}
= q^*
\text{,}
\label{eq:modified-problem}
\end{equation}
subject to the homogeneous boundary conditions
\begin{subequations}
	\begin{align}
	a_\pm \tilde{T} + b_\pm \tilde{T} \frac{\partial}{\partial \hat{r}} \tilde{T} = 0 \text{,}
	\label{eq:BChomoga}\\
	c_\pm \tilde{T} + d_\pm \tilde{T} \frac{\partial}{\partial \hat{z}} \tilde{T}  = 0 \text{,}
	\label{eq:BChomogb}
	\end{align}
\end{subequations}
where 
\begin{equation}
q^* = q - \left(
- \alpha^2 k_{r}\frac{\partial^{2}T_e}{\partial \hat{r}^{2}}
- \frac{\alpha^2 k_r}{1+r_{in} + \hat{r}} \frac{\partial T_e}{\partial \hat{r}}
- \beta^2 k_{z}\frac{\partial^{2}T_e}{\partial \hat{z}^{2}}
\right)
\text{.}
\end{equation}

The boundary lifting is similar to that in \cite{Doha2012a}, except that here we apply it to cylindrical coordinates and neglect the corner component of the lifting (since we are assuming constant external temperatures).

\subsection{Chebyshev-Galerkin approximation\label{sec:SG_Galerkin}}
Now we would like to solve the modified problem (\ref{eq:modified-problem}) using the Galerkin method. The first step is to multiply the equation by a test function, $\nu$, and integrate over the entire domain,

\begin{multline}
\left( 
\frac{1+\alpha r_{in} + \hat{r}}{\alpha}
\left[ \rho c_{p}\frac{\partial \tilde{T}}{\partial t}
- \alpha^2 k_{r}\frac{\partial^{2}\tilde{T}}{\partial \hat{r}^{2}}
- \frac{\alpha^2 k_r}{1+\alpha r_{in} + \hat{r}} \frac{\partial \tilde{T}}{\partial \hat{r}}
\right.
\right.
\\
\left.
\left.
- \beta^2 k_{z}\frac{\partial^{2}\tilde{T}}{\partial \hat{z}^{2}} \right] , \,\nu \right)
= \left( \frac{1+\alpha r_{in} + \hat{r}}{\alpha} \, q^* , \,\nu \right)
\label{eq:galerkin}
\end{multline}
where $\left( {f} , \, \nu \right)$ denotes the integral of ${f}$ weighted by $\nu$, throughout the domain $\hat{r},\hat{z}$,
\begin{equation}
({f}, \, \nu) = \int\limits_{-1}^{1} \int\limits_{-1}^{1} {f}(\hat{r},\hat{z}) \nu(\hat{r},\hat{z}) \mathrm{d}\hat{r} \mathrm{d}\hat{z}
\text{.}
\label{eq:integral}
\end{equation}
Note the inclusion of the $r$ terms as defined in (\ref{eq:scale_r2}) on each side of eq. (\ref{eq:galerkin}) to account for cylindrical coordinates.

The second step is to approximate the solution $\tilde{T}(\hat{r},\hat{z},t)$ with a finite number of functions
\begin{equation}
\tilde{T} = \sum\limits_{k=0}^{N} \sum\limits_{j=0}^{N} x_{kj} \phi^{\hat{r}}_k(\hat{r},a_\pm,b_\pm) \phi^{\hat{z}}_j(\hat{z},c_\pm,d_\pm)
\end{equation}
where $x_{kj}$ are unknown solution coefficients, and $\phi^{\hat{r}}_k$ and $\phi^{\hat{z}}_j$ are the basis functions which must satisfy the homogeneous boundary conditions (\ref{eq:BChomoga}) and (\ref{eq:BChomogb}) respectively. For simplicity, the same number of basis functions, $N$, are chosen for the radial and axial directions but it should be noted that different values could be chosen if one dimension required greater resolution than the other.

Suitable basis functions can be found as follows.
First we find basis functions in the radial and axial directions separately.
Each of these is obtained in a similar manner to that of the toy problem (eqs.~\ref{eq:toy_basis_1} - \ref{eq:toy_basis_2}):
we denote by $C_k$ the $k$th degree Chebyshev polynomial of the first kind and then let $\{\phi^{\hat{r}}_k(\hat{r},a_\pm,b_\pm)\}_{k=0}^N $ be a basis in the radial direction such that 
\begin{equation}
\phi^{\hat{r}}_k = C_k(\hat{r}) + \zeta_k^{\hat{r}} C_{k+1}(\hat{r}) + \eta_k^{\hat{r}} C_{k+2}(\hat{r}) \text{,}
\end{equation}
where $\zeta_k^{\hat{r}}(a_\pm,b_{\pm})$ and $\eta_k^{\hat{r}}(a_\pm,b_\pm)$ are defined as before such that they adhere to the boundary conditions in the radial direction.

Similarly, we let $\{\phi^{\hat{z}}_j(\hat{z},c_\pm,d_\pm)\}_{j=0}^N $ be a basis in the axial direction such that 
\begin{equation}
\phi^{\hat{z}}_j = C_j(\hat{z}) + \zeta_j^{\hat{z}} C_{j+1}(\hat{z}) + \eta_j^{\hat{z}} C_{j+2}(\hat{z}) \text{,}
\end{equation}
where $\zeta_j^{\hat{z}}(c_\pm,d_\pm)$ and $\eta_j^{\hat{z}}(c_\pm,d_\pm)$ adhere to the boundary conditions in the axial direction.

As mentioned previously the choice of basis function is not limited to Chebyshev polynomials: in general, various Jacobi polynomials may work equally well. In this case, we verified that almost identical results were obtained using Legendre polynomials.

\subsection{Boundary lifting function\label{sec:SG_Boundary_lifting}}

We now wish to determine the boundary lifting function, $T_e(\hat{r},\hat{z})$.
Recall that $T_e(\hat{r},\hat{z})$ must satisfy the original (non-homogeneous) boundary conditions (\ref{eq:BC1b}), such that when subtracted from the actual temperature, $T$, (eq.~\ref{eq:boundary_auxiliary}) the resulting auxiliary function, $\tilde{T}$, satisfies the homogeneous boundary conditions.
Hence, the aim of this section is to derive a function which is constant with respect to time and satisfies the non-homogeneous boundary conditions.
Since this cannot be achieved exactly, the boundary conditions must instead be satisfied in a weak sense (i.e. the solution converges as more basis functions are included).
To begin, we adopt a similar approach to \cite{Doha2012a} by assuming a form of the solution as follows:
\begin{equation}
T_e = \left[ \sum\limits_{k=0}^{N}  \left( d_k^{\mathrm{I}} \hat{z} +d_k^{\mathrm{II}} \hat{z}^2 \right) \phi^{\hat{r}}_k \right]
+
\left[ \sum\limits_{j=0}^{N}  \left( d_j^{\mathrm{III}} \hat{r} + d_j^{\mathrm{IV}} \hat{r}^2 \right) \phi^{\hat{z}}_j \right]
\text{.}
\label{eq:approx-function}
\end{equation}
The conditions at the vertical and horizontal sides are defined as follows. For the right side, we have
\begin{equation}
c_+ T_e + d_+ \frac{\partial T_e}{\partial \hat{z}} \approx e_{r} 
\text{\quad at $\hat{z} = +1$}
\label{eq:weak1}
\end{equation}
where $\approx$ denotes weakly satisfying conditions. Substituting (\ref{eq:approx-function}) in (\ref{eq:weak1}) gives
\begin{equation}
\begin{split}
\sum\limits_{k = 0}^{N}
\left( d_k^{\mathrm{I}}(c_+ \hat{z} + d_+)  + d_k^{\mathrm{II}}(c_+ \hat{z}^2 + 2 d_+ \hat{z}) \right)
\phi^{\hat{r}}_k & \\
+ \sum\limits_{k = 0}^{N}
\left( \hat{r} d_j^{\mathrm{III}} + \hat{r}^2 d_j^{\mathrm{IV}} \right)
\left( c_+ \phi^{\hat{z}}_j + d_+ \frac{\partial \phi^{\hat{z}}_j}{\partial \hat{z}} \right)
& \approx e_{r}
\text{.}
\end{split}
\label{eq:weak2}
\end{equation}
Substituting for $\hat{z}=1$ and noting that the term $\left( c_+ \phi^{\hat{z}}_j + d_+ \frac{\partial \phi^{\hat{z}}_j}{\partial z} \right)$ is equal to zero (since the basis functions by definition satisfy the homogeneous boundary conditions), eq. (\ref{eq:weak2}) simplifies to
\begin{equation}
\sum\limits_{k = 0}^{N} \left( d_k^{\mathrm{I}}(c_+ + d_+)  + d_k^{\mathrm{II}}(c_+ + 2 d_+) \right)
\phi^{\hat{r}}_k \approx e_{r}
\text{.}
\end{equation}
The weak boundary condition can now be replaced by the appropriate integral equation,
\begin{equation}
\sum\limits_{k = 0}^{N} \left( d_k^{\mathrm{I}}(c_+ + d_+)  + d_k^{\mathrm{II}}(c_+ + 2 d_+) \right)
\left<\phi^{\hat{r}}_k, \, \phi^{\hat{r}}_i\right>_{\hat{r}} = \left<e_{r}, \, \phi^{\hat{r}}_i\right>_{\hat{r}}
\text{,}
\label{eq:weak3a}
\end{equation}
where $\left< {f} , \, {g} \right>_{\hat{r}}$ denotes the integral,
\begin{equation}
\left<{f}, \, {g} \right>_{\hat{r}} = \int\limits_{-1}^{1}
\left(\frac{1+\hat{r}+\alpha r_{in}}{\alpha}\right)
{f}(\hat{r}) {g} (\hat{r}) \mathrm{d}\hat{r}
\text{.}
\end{equation}
Through a similar process for the left, top and bottom sides, we find that
\begin{equation}
\sum\limits_{k = 0}^{N} \left( d_k^{\mathrm{I}}(-c_- + d_-)  + d_k^{\mathrm{II}}(c_- - 2 d_-) \right)
\left<\phi^{\hat{r}}_k, \, \phi^{\hat{r}}_i\right>_{\hat{r}} = \left<e_{l}, \, \phi^{\hat{r}}_i\right>_{\hat{r}}
\text{,}
\label{eq:weak3b}
\end{equation}
\begin{equation}
\sum\limits_{j = 0}^{N} \left( d_j^{\mathrm{III}}(a_+ + b_+)  + d_j^{\mathrm{IV}}(a_+ + 2 b_+) \right)
\left<\phi^{\hat{z}}_j, \, \phi^{\hat{z}}_i\right>_{\hat{z}} = \left<e_{t}, \, \phi^{\hat{z}}_i\right>_{\hat{z}}
\text{,}
\label{eq:weak3c}
\end{equation}
\begin{equation}
\sum\limits_{j = 0}^{N} \left( d_j^{\mathrm{III}}(-a_- + b_-)  + d_j^{\mathrm{IV}}(a_- - 2 b_-) \right)
\left<\phi^{\hat{z}}_j, \, \phi^{\hat{z}}_i\right>_{\hat{z}} = \left<e_{b}, \, \phi^{\hat{z}}_i\right>_{\hat{z}}
\text{,}
\label{eq:weak3d}
\end{equation}
where $\left< {f} , \, {g} \right>_{\hat{z}}$ denotes the integral,
\begin{equation}
\left<{f}, \, {g} \right>_{\hat{z}} = \int\limits_{-1}^{1} {f}(\hat{z}) {g}(\hat{z}) \mathrm{d}\hat{z}
\text{.}
\end{equation}
Equations (\ref{eq:weak3a}) and (\ref{eq:weak3b}) and equations (\ref{eq:weak3c}) and (\ref{eq:weak3d}) form two linear systems which can be each solved for the corresponding $\mathrm{d}^\sigma$ vectors:
\begin{subequations}
	\begin{equation}
	\mathbf{d}^{\mathrm{I}} = \frac{k_4 e_{r} - k_2 e_{l}}{k_1 k_4 - k_2 k_3}
	(\mathbf{P}^{rl})^{-1} \mathbf{s}^{rl}
	\text{,}
	\label{eq:lin-sys-a}
	\end{equation}
	\begin{equation}
	\mathbf{d}^{\mathrm{II}} = \frac{k_1 e_l - k_3 e_r}{k_1 k_4 - k_2 k_3}
	(\mathbf{P}^{rl})^{-1} \mathbf{s}^{rl}
	\text{,}
	\end{equation}
	\begin{equation}
	\mathbf{d}^{\mathrm{III}} = \frac{j_4 e_{t} - j_2 e_{b}}{j_1 j_4 - j_2 j_3}
	(\mathbf{P}^{tb})^{-1}
	\mathbf{s}^{tb}
	\text{,}
	\end{equation}
	\begin{equation}
	\mathbf{d}^{\mathrm{IV}} = \frac{j_1 e_{b} - j_3 e_{t}}{j_1 j_4 - j_2 j_3}
	(\mathbf{P}^{tb})^{-1}
	\mathbf{s}^{tb}
	\text{,}
	\label{eq:lin-sys-b}
	\end{equation}
\end{subequations}
where $\{\mathbf{d}^{\mathrm{\sigma}}  = (d_0^{\sigma}, d_1^{\sigma}, ... , d_2^{\sigma})^T; \sigma = \mathrm{I}, \mathrm{II}, \mathrm{III} \text{ and } \mathrm{IV} \}$ are vectors of unknown expansion coefficient, $\{\mathbf{s}^{\mathrm{\sigma}}  = (s_0^{\sigma}, s_1^{\sigma}, ... , s_N^{\sigma})^T; \sigma = rl, \text{ and } tb \}$ are vectors of known source terms, and
\begin{align}
\begin{split}
k_1 & = c_+ + d_+ ,\qquad  		j_1  = a_+ + b_+ ,\\
k_2 & = c_+ + 2 d_+ ,\qquad 	j_2  = a_+ + 2 b_+ ,\\
k_3 & = d_- - c_- ,\qquad 		j_3  = b_- - a_- ,\\
k_4 & = c_- - 2 d_- ,\qquad 	j_4  = a_- - 2 b_-.
\end{split}
\end{align}
The $\mathbf{P}$ matrix for the right and left edges is defined as $\{\mathbf{P}^{rl}  = p^{rl}_{i,k}; \, i, k = 0, 1,..., N \}$, where
\begin{subequations}
	\begin{equation}
	p^{rl}_{i,k} = 
	\int\limits_{-1}^{1} \left(\frac{1+\hat{r}+\alpha r_{in}}{\alpha}\right) \phi^{\hat{r}}_i \phi^{\hat{r}}_k \mathrm{d}\hat{r},
	\end{equation}
	and that for top and bottom sides as $\{\mathbf{P}^{tb}  = p^{tb}_{i,j}; \, i, j = 0, 1,..., N \}$, where
	\begin{equation}
	p^{tb}_{i,j} =
	\int\limits_{-1}^{1} \phi^{\hat{z}}_i \phi^{\hat{z}}_j \mathrm{d}\hat{z}.
	\end{equation}
\end{subequations}
Lastly, the source terms are defined as $\{\mathbf{s}^{\mathrm{\sigma}}  = s^\sigma_{i}; \, i = 0, 1,..., N; \, \sigma = rl \text{ and } tb \}$, where
\begin{subequations}
	\begin{equation}
	s^{rl}_{i} =
	\int\limits_{-1}^{1} \left(\frac{1+\hat{r}+\alpha r_{in}}{\alpha}\right) \phi^{\hat{r}}_i \mathrm{d}\hat{r},
	\end{equation}
	\begin{equation}
	s^{tb}_{i} =
	\int\limits_{-1}^{1} \phi^{\hat{z}}_i \mathrm{d}\hat{z}.
	\end{equation}
\end{subequations}
Thus, an explicit expression for the unknown expansion coefficients, $\mathbf{d}^\sigma$ is obtained, and so the boundary lifting function, $T_e$, can be considered known at this stage.

\subsection{Solution algorithm\label{sec:SG_Sol}}

Finally, we present the solution algorithm to the modified problem. Equation (\ref{eq:modified-problem}) can be expressed in state space form, as follows:
\begin{equation}
\mathbf{E}\mathbf{\dot{x}} = \mathbf{A}\mathbf{x} + \mathbf{B}\mathbf{u}
\label{eq:State-space-model-matrices-2D}
\end{equation}
where
\begin{align}
\begin{split}
\mathbf{x} = \left(
x_{00},
x_{10},
...
x_{N0},
x_{01},
x_{11},
...,
x_{N1},
...,
x_{0N},
...,
x_{NN}
\right)^T
\text{,}
\end{split}
\label{eq:SG_states}
\end{align}
and $\mathbf{u} = \left[q(t), \, 1\right]^T$.
The system matrices are defined as follows. First, let us denote the column vector
\begin{equation}
\begin{split}
{\Psi} = \left(
\phi^{\hat{r}}_0 \phi^{\hat{z}}_0,
\phi^{\hat{r}}_1 \phi^{\hat{z}}_0,
...,
\phi^{\hat{r}}_N \phi^{\hat{z}}_0,
\phi^{\hat{r}}_0 \phi^{\hat{z}}_1,
\phi^{\hat{r}}_1 \phi^{\hat{z}}_1,
...,
\right.
\\
\left.
\phi^{\hat{r}}_N \phi^{\hat{z}}_1,
...,
\phi^{\hat{r}}_0 \phi^{\hat{z}}_N,	
...,
\phi^{\hat{r}}_N \phi^{\hat{z}}_N
\right)^T
\end{split}
\end{equation}
and take $\nu = \psi_i$, where $\psi_i$ is the $i$th element of the vector $\Psi$; then:
\begin{equation}
\mathbf{E}(i,j) = \rho c_p \left( \frac{1+\hat{r}+\alpha r_{in}}{\alpha} \psi_j, \psi_i \right),
\label{eq:SG2-eq1}
\end{equation}
where $\mathbf{E}(i,j)$ denotes the element in the $i$th row and $j$th column of the matrix $\mathbf{E}$,
\begin{align}
& \mathbf{A}(i,j) =
\\
& \left(
\frac{1+\hat{r}+\alpha r_{in}}{\alpha}
\left[
\alpha^2 k_r \frac{\partial^2 \psi_j}{\partial \hat{r}^2}
+ \beta^2 k_z \frac{\partial^2 \psi_j}{\partial \hat{z}^2}
\right]
+ \alpha k_r \frac{\partial \psi_j}{\partial \hat{r}}
,
\psi_i
\right)
\end{align}
for $i,j = 0,1,...,N$, and, 
\begin{align}
& \mathbf{B}(i,1) = \left( \frac{1+\hat{r}+\alpha r_{in}}{\alpha} , \psi_{i} \right) \text{, }
\\ \nonumber &
\mathbf{B}(i,2) = 
\\ \nonumber &
\left(
\frac{1+\hat{r}+\alpha r_{in}}{\alpha}
\left[
\alpha^2 k_r \frac{\partial^2 T_{e}}{\partial \hat{r}^2}
+ \beta^2 k_z \frac{\partial^2 T_{e}}{\partial \hat{z}^2}
\right]
+ \alpha k_r \frac{\partial T_{e}}{\partial \hat{r}}
,
\psi_i
\right)
\end{align}
for $i = 0,1,...,N$.

The complete solution to the original non-homogeneous problem is then given by:
\begin{equation}
T(\hat{r},\hat{z},t) = \tilde{T}(\hat{r},\hat{z},t) + T_e(\hat{r},\hat{z})
\end{equation}
We choose as outputs the temperatures at the bottom-centre, left-centre, top-centre and right-centre locations, i.e. $T_{1} = T(\hat{r}=-1,\hat{z}=0)$, $T_{2} = T(\hat{r}=0,\hat{z}=-1)$, $T_{3} = T(\hat{r}=1,\hat{z}=0)$ and $T_{4} = T(\hat{r}=0,\hat{z}=1)$ (see Figure \ref{fig:model-outputs}).
Thus, the outputs are given by:
\begin{equation}
\mathbf{y} = \mathbf{C}\mathbf{x} + \mathbf{T_e}
\label{eq:model-outputs}
\end{equation}
where $\mathbf{y} = \left( T_{1}, T_{2}, T_{3}, T_{4} \right)^T $, $\mathbf{T_e} = \left( T_{e, 1}, T_{e, 2}, T_{e, 3}, T_{e, 4} \right)^T $, and,
\begin{equation}
\mathbf{C} := \; 
\begin{cases}
\mathbf{C}({1,j}), & = \psi_j(\hat{r}=-1,\hat{z}=0),\\
\mathbf{C}({2,j}), & = \psi_j(\hat{r}=0,\hat{z}=-1),\\
\mathbf{C}({3,j}), & = \psi_j(\hat{r}=1,\hat{z}=0),\\
\mathbf{C}({4,j}), & = \psi_j(\hat{r}=0,\hat{z}=1)
\end{cases}
\label{eq:SG2-eq2}
\end{equation}
for $j = 0,1,...,N$.

Note that the mean temperature, $\overline{T}$, may also included as an output ($\overline{T}$ is used in Part II of this paper for computing the overall cell electrochemical impedance).
Hence, an additional row is appended to the $\mathbf{C}$ matrix,
\begin{equation}
\mathbf{C}(5,j) = \frac{1}{H} \frac{2}{r_{out}^2 - r_{in}^2} \int\limits_{0}^{H}  \int\limits_{r_{in}}^{r_{out}} r \psi_j(r,z) \mathrm{d}r \mathrm{d}z,
\end{equation}
for $j = 0,1,...,N$, which in the scaled coordinates becomes
\begin{equation}
\mathbf{C}(5,j) = \frac{1}{\alpha \beta} \int\limits_{-1}^{1}  \int\limits_{-1}^{1}
\frac{1+\hat{r}+\alpha r_{in}}{\alpha}
\psi_j(\hat{r},\hat{z}) \mathrm{d}\hat{r} \mathrm{d}\hat{z}.
\end{equation}
An additional element must also be included in the boundary lifting function,
\begin{equation}
T_{e,5} = \frac{1}{H} \frac{2}{r_{out}^2 - r_{in}^2} \int\limits_{0}^{H}  \int\limits_{r_{in}}^{r_{out}} r T_e(r,z) \mathrm{d}r \mathrm{d}z,
\end{equation}
which in the scaled domain becomes
\begin{equation}
T_{e,5} = \frac{1}{\alpha \beta} \int\limits_{-1}^{1}  \int\limits_{-1}^{1}
\frac{1+\hat{r}+\alpha r_{in}}{\alpha}
T_e (\hat{r},\hat{z}) \mathrm{d}\hat{r} \mathrm{d}\hat{z}.
\end{equation}
With these equations, the mean temperature is computed as the fifth output.

The frequency domain response of the above linear system, ${H}(s)$, is calculated by
\begin{equation}
\mathbf{H}(s) = {\mathbf{C}}(s\mathbf{I}-{\mathbf{A}})^{-1}{\mathbf{B}}
\label{eq:freq-resp}
\end{equation}
where $s=j\omega$ is the Laplace variable and $\mathbf{I}$ is the identity matrix.

The above algorithm was implemented both numerically (using clenshaw-curtis quadrature) and analytically using the Matlab Symbolic Maths Toolbox. The numerical implementation allows the state matrices to be generated more efficiently than the symbolic approach, although the resulting state space model is identical (and therefore equally efficient) in each case.

\section{Results and discussion\label{sec:SG_Results}}

To validate the SG model, the results were compared with high fidelity FEM simulations,  implemented using the Matlab Partial Differential Equation Toolbox. To ensure the accuracy of the FEM solution, a fine mesh consisting of 3,760 elements was used. The time step for both the SG and FEM models was set at 1 s.

The thermo-physical parameters chosen for the model validation are shown in Table \ref{table-1}. The dimensions were chosen to match those of the large format lithium-ion cell employed in \cite{Roscher2015}, and the remaining thermal parameters were chosen based on typical properties of lithium iron phosphate cells (\cite{Muratori2010a, fleckenstein2013thermal, Kim2014b}).

\begin{table}[hbt]
	\centering
	\caption{Thermophysical properties for model validation.}
	\label{table-1}
	\begin{tabular}{@{}llll@{}}
		\toprule
		Parameter & Symbol & Value &  \\ \midrule
		Inner radius & $r_{in}$ & $4$ mm &  \\
		Outer radius & $r_{out}$ & $32$ mm &  \\
		Height & $H$ & $198$ mm &  \\
		Density & $\rho$ & $2,118$ kg m\textsuperscript{-3} &  \\
		Specific heat capacity & $c_p$ & $765$ J kg\textsuperscript{-1} K\textsuperscript{-1} &  \\
		Radial thermal conductivity & $k_r$ & $0.66$ W mK\textsuperscript{-1} &  \\
		Axial thermal conductivity & $k_z$ & $66$ W mK\textsuperscript{-1} &  \\ \bottomrule
	\end{tabular}
\end{table}

\subsection{Time domain}

Time domain simulations were carried out using two different cooling scenarios, as shown in Table \ref{table-2}. Case 1 represents forced convection air cooling, with equal temperatures and convection coefficients at each of the external sides. Case 2 represents forced convection liquid cooling at the left end of the cell (for instance, via a cooling plate \cite{Jarrett2011a}), and {mild forced convection} to the ambient air at the remaining faces. Thus, the temperature at the left face is set to $3 ^\circ$C and a large convection coefficient typical of forced cooling via water or glycol is applied, whereas the remaining faces are exposed to a small convection coefficient at $18 ^\circ$C. This case was chosen to highlight the ability of the model to account for different external temperatures and/or convection coefficients at each side. Note that the convection coefficient at the bottom side (the inner radius of the jelly roll) was set to zero in both cases since negligible cooling occurs here in a typical thermal management system; however a non-zero value could easily be applied if it were required.

\begin{table}[hbt]
	\centering
	\caption{Convection coefficients for the two cooling scenarios.}
	\label{table-2}
	\begin{tabular}{@{}llllll@{}}
		\toprule
		& \multicolumn{2}{c}{Case 1} & \multicolumn{2}{c}{Case 2} &   \\
		& $h$ (W m\textsuperscript{-2}) & $T_\infty$ ({$^\circ$}C) & $h$ (W m\textsuperscript{-2}) & $T_\infty$ ({$^\circ$}C) &  \\
		\midrule
		Left & $100$  & $18$  & $400$  & $3$  & \\
		Right & $100$  & $18$  &  $30$  & $18$  &  \\
		Top & $100$ & $18$  &  $30$ & $18$  &  \\
		Bottom & $0$ & $18$  & $0$ & $18$  &  \\
		\bottomrule
	\end{tabular}
\end{table}

\begin{figure}
	\centering
	\includegraphics[width=0.85\columnwidth]{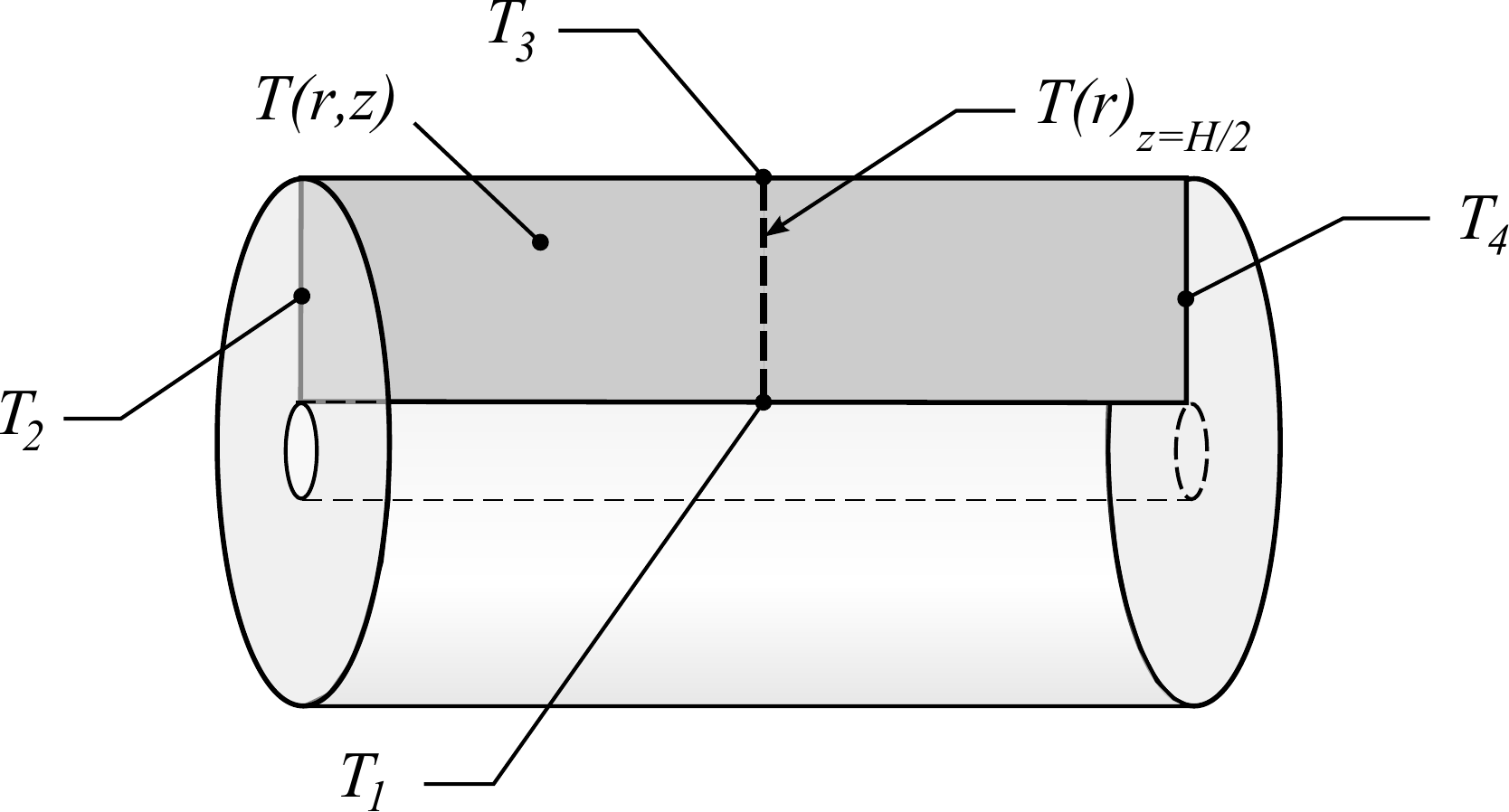}
	\caption[Schematic of cell showing the model outputs]{Schematic of cell showing the model outputs displayed in Figures \ref{fig:time-resp-air} and \ref{fig:time-resp-plate}.}
	\label{fig:model-outputs}
\end{figure}

\begin{figure*}[!hbt]
	\centering
	\includegraphics[width=0.99\textwidth]{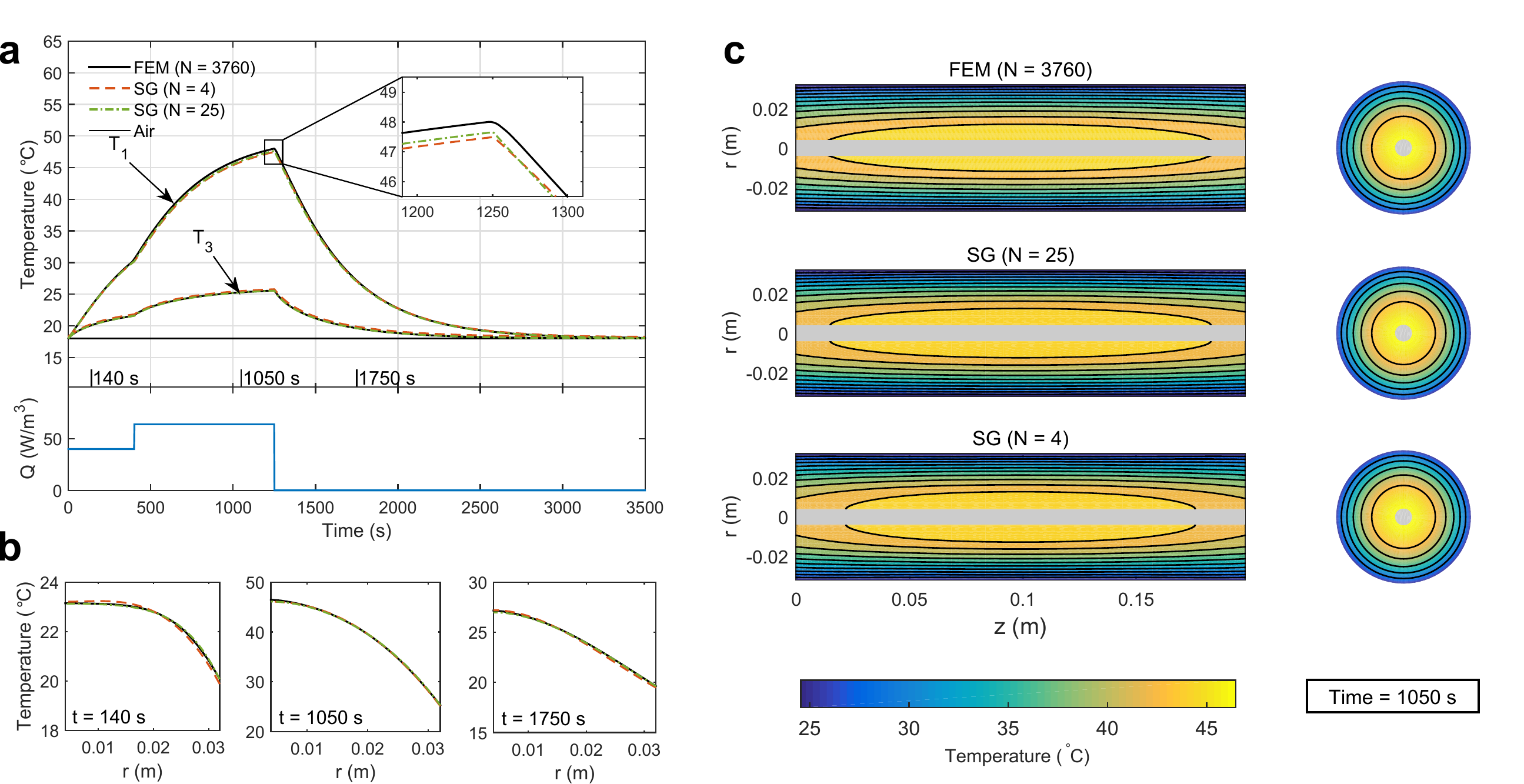}
	\caption[Comparison of SG vs. FEM results for Case 1: forced air convection]{Comparison of SG vs. FEM results for Case 1: forced air convection.  (a) Evolution of temperatures, $T_{1}$ (core) and $T_{3}$ (outer surface), with pulsed load profile shown in subplot; (b) temperature distribution along the centre-line ($z=H/2$) of the cell at denoted times.}
	\label{fig:time-resp-air}
\end{figure*}

\begin{figure*}[!hbt]
	\centering
	\includegraphics[width=0.99\textwidth]{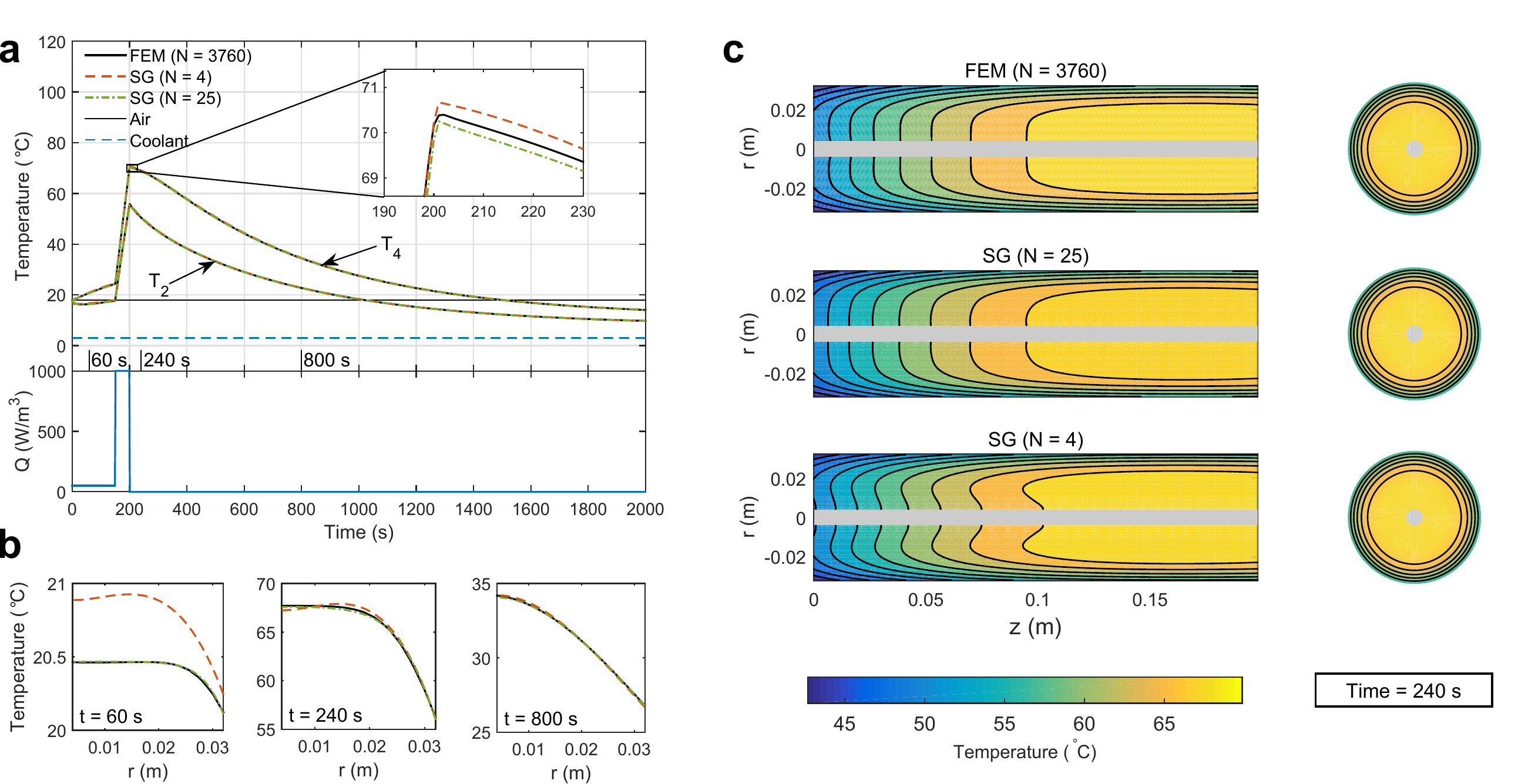}
	\caption[Comparison of SG vs. FEM results for Case 2: end-plate liquid cooling]{Comparison of SG vs. FEM results for Case 2: end-plate liquid cooling.  (a) Evolution of temperatures, $T_{2}$ and $T_{4}$, with pulsed load profile shown in subplot; (b) temperature distribution along the centre-line ($z=H/2$) of the cell at denoted times}
	\label{fig:time-resp-plate}
\end{figure*}

To ensure highly transient conditions with large internal temperature gradients, pulsed power load profiles with large heat generation rates but relatively short durations were applied. The load profiles and resulting temperature distributions for the two cases are compared against the corresponding FEM solutions in Figures \ref{fig:time-resp-air} and \ref{fig:time-resp-plate}. The locations of the model outputs are shown in Figure \ref{fig:model-outputs}.

The results show that the SG method with $N_s = 4$ states (i.e. 2 basis functions in each of the radial and axial directions) is capable of accurately capturing the temperatures in both cases. Slightly greater accuracy is achieved using $N_s = 9$ states (3 basis functions in each direction). In Case 1, the temperatures, $T_{1}$ and $T_{3}$ (i.e. the first and third model outputs from eq.~(\ref{eq:model-outputs}))  are plotted (see Figure \ref{fig:time-resp-air}a) since the temperature gradient is primarily in the radial direction in this case. In Case 2, the temperatures $T_{2}$ and $T_{4}$ were plotted since a large gradient occurs in the axial direction in this case. Plots of $T(r)_{z=H/2}$ are shown in part (b) of these figures. In each case, the first two sub-plots show the distribution at instances of highly transient dynamics (i.e. just after a load is applied or removed), which is more difficult to capture accurately for a low-order thermal model, although the results are still in good agreement in both cases. The third sub-plot in each case shows the solution at a later time - when the problem is dominated by slow thermal dynamics - and the SG and FEM solutions are in nearly perfect agreement at these times. The full 2-D contour plots are shown in part (c) of each figure. Again, the solution is plotted at an instant with transient dynamics to demonstrate the ability of the low-order model to accurately simulate the temperature field under these conditions. The SG model is in good agreement with the FEM solution, although the advantage of increasing the number of model states is apparent in Figure \ref{fig:time-resp-plate}(c), as the result using $N_s = 9$ states is in better agreement with that of the FEM solution than the result with $N_s = 4$ states.

Lastly, we note that the model with 4 states is of a similar order to an equivalent circuit thermal model and so could be applied with similar computational efficiency. Moreover, a similar order model implemented using a spectral-collocation method would not give results as accurate as those presented in this section, since in the SG method the boundary conditions are implicitly satisfied by the basis functions, whereas additional equations are required to enforce the boundary conditions in the collocation case.

\subsection{Frequency domain}
In this section we compare the frequency response of the low-order SG models against a baseline solution obtained by using an SG model with a large number of states ($N_s = 225$). Specifically, we examine the impact of changes in heat generation on $T_{1}$, given by the transfer function $H(s)=T_{1}(s)/q(s)$. This is calculated using eq.~(\ref{eq:model-outputs}), using only the rows and columns of the $\mathbf{B}$ and $\mathbf{C}$ matrices corresponding to the heat generation input and $T_{1}$ output.

\begin{figure}[hbt]
	\centering
	\includegraphics[width=1\columnwidth]{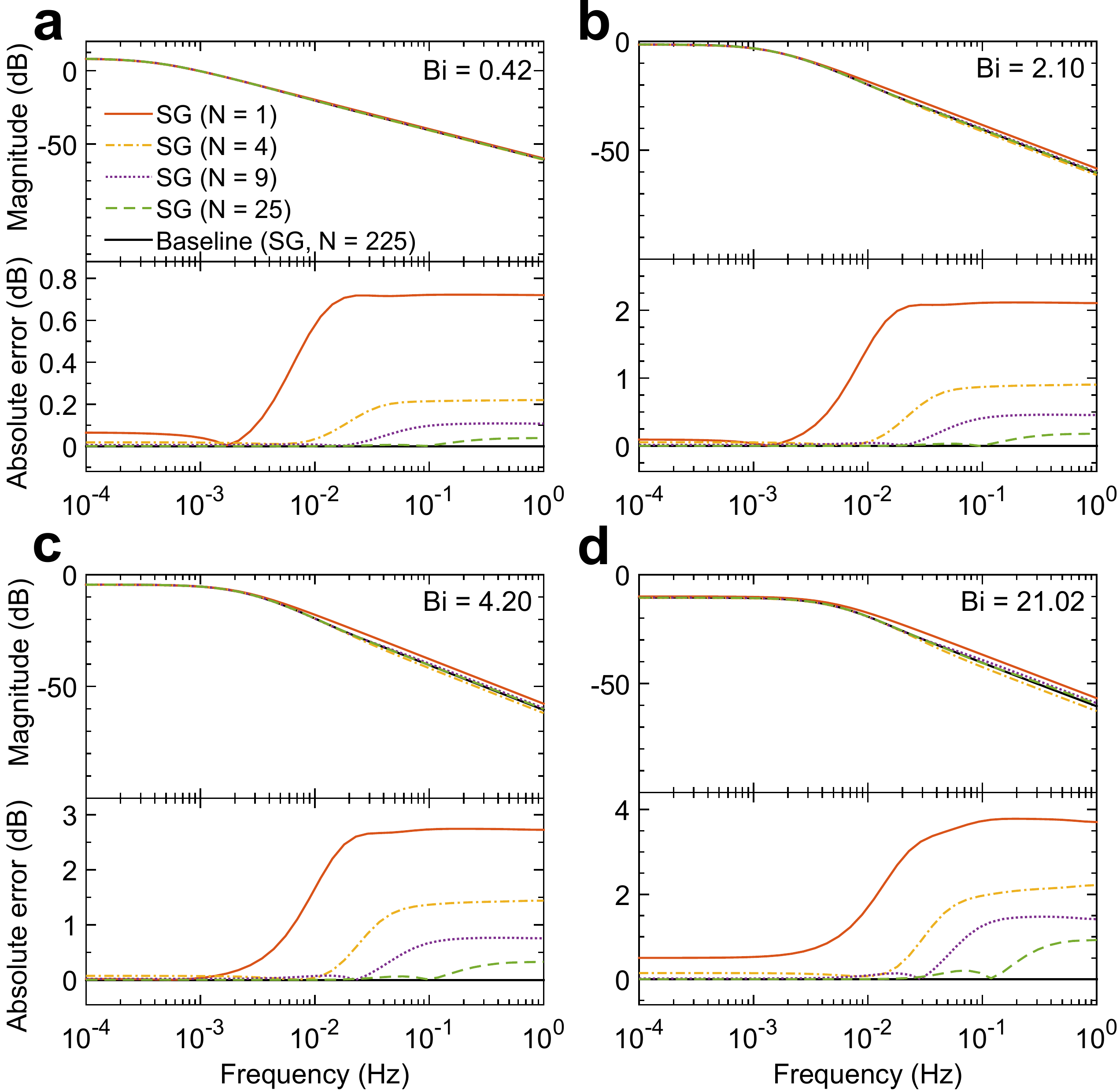}
	\par
	\caption[Frequency response of the SG method for a range of model orders and Biot numbers]{Frequency response, $T_{1}(s)/q(s)$, of the SG method with different model orders for a range of Biot number conditions as indicated. A large order ($N_s=225$) SG method is used as the baseline solution.}
	\label{fig:freq-resp}
\end{figure}

The parameters chosen are the same as those in Case 1 of the previous section. However, a variety of different Biot numbers ($Bi = h (r_{out}-r_{in})/k_{\hat{r}} $) are obtained by varying the magnitude of the convection coefficients on each side. Convection coefficients of $h = \{10$, $50$, $100$, $500\}$ $\mathrm{Wm}^{-2}$ are chosen, resulting in Biot numbers of $Bi = \{0.42$, $2.10$, $4.20$ and $21.02\}$, respectively. The size of the model is varied by increasing the number of basis functions in both the radial and axial directions. Models with 1, 2, 3 and 5 basis functions in each direction are chosen, resulting in $N_s =$ 1, 4, 9 and $25$ states respectively.

Figures~\ref{fig:freq-resp}(a)-(d) show the magnitude of the frequency response in the range $f = 1 \times 10^{-4}$ to $1 \times 10^{0}$ for each of the four Biot numbers, along with the error of the low order models relative to the high fidelity solution.
These plots show that as the model order (i.e. the number of states) is increased, the magnitude of the error is reduced.  Moreover, for all cases, there is a critical frequency at which the error becomes non-negligible, and this critical frequency increases as the model order is increased.
We also note that the error increases as (i) the Biot number increases, and (ii) the perturbation frequency increases. These trends are as expected.
Thus, if a particular application involves larger Biot numbers or higher frequencies (for instance, due to larger cells, more aggressive drive cycles or higher performance cooling), the model order could be increased accordingly to achieve a required accuracy.

\section{Conclusions\label{sec:Conclusion}}

Computationally efficient thermal models are necessary for control-oriented applications.
The model presented in this paper is of a similar order to a thermal equivalent circuit (TEC) model - and so could be applied with similar computational efficiency - but has much greater spatial resolution.
It could therefore provide greater accuracy than a TEC if used as part of a state-estimation scheme.
Morevover, although we have presented here a model for a cylindrical cell, it can easily be modified to apply to 2-D simulation of prismatic cells.
However, it is difficult to apply the SG method to more involved problems, due to the complexity of it's implementation.
Hence, for configurations involving several cells or non-uniform cooling, TECs may remain favourable due to their relative simplicity.

In the companion contribution \cite{richardson2016on}, the model is incorporated into a state estimation scheme and the predicted temperatures at four locations (one internal and three on the cell surface) are validated experimentally against thermocouple measurements.


\section*{Acknowledgements}
This work was funded by a NUI Travelling Scholarship, a UK EPSRC Doctoral Training Award,
 the Foley-Bejar scholarship from Balliol College, University of Oxford, and the RCUK Energy Programmes's STABLE-NET project (ref. EP/L014343/1).

\section*{Nomenclature}

 \begin{supertabular}{ll}
 	$\mathbf{A}$, $\mathbf{B}$, $\mathbf{C}$, $\mathbf{D}$, $\mathbf{E}$ & system matrices\tabularnewline
 	$a_{i}$ & polynomial coefficients\tabularnewline
 	$c_{p}$ & specific heat capacity {[}J kg\textsuperscript{-1} K\textsuperscript{-1}{]}\tabularnewline
 	$C_k$ & $k$th deg. Chebyshev poly. of 1\textsuperscript{1st} kind\tabularnewline
 	$h$ & convection coefficient {[}W m\textsuperscript{-2}{]}\tabularnewline
 	$H$ & cell height\tabularnewline
 	$\mathbf{H}(s)$ & system transfer function\tabularnewline
 	$\mathbf{I}$ & identity matrix\tabularnewline
 	$j$ & imaginary number\tabularnewline
 	$k_{r}$ & radial thermal conductivity {[}W m\textsuperscript{-1} K\textsuperscript{-1}{]}\tabularnewline
  	$k_{z}$ & axial thermal conductivity {[}W m\textsuperscript{-1} K\textsuperscript{-1}{]}\tabularnewline
 	$q$ & volumetric heat generation {[}W{]}\tabularnewline
 	$N_s$ & number of model states\tabularnewline
 	$r$ & radial coordinate {[}m{]}\tabularnewline
 	$\hat{r}$ & scaled radial coordinate {[}m{]}\tabularnewline
	$r_{in}$ & inner radius {[}m{]}\tabularnewline
	$r_{out}$ & outer radius {[}m{]}\tabularnewline
 	$s$ & Laplace variable\tabularnewline
 	$T$ & temperature {[}K{]}\tabularnewline
	$\tilde{T}$ & auxiliary temperature function {[}K{]}\tabularnewline
	$T_e$ & boundary-lifting function {[}K{]}\tabularnewline
 	$t$ & time {[}s{]}\tabularnewline
 	$\mathbf{x}$ & system state\tabularnewline
 	$\mathbf{y}$ & system output\tabularnewline
 	$z$ & axial coordinate {[}m{]}\tabularnewline
 	$\hat{z}$ & scaled axial coordinate {[}m{]}\tabularnewline
 	& \tabularnewline
 	\multicolumn{2}{l}{\emph{Abbreviations}}\tabularnewline
 	BVP & Boundary Value Problem\tabularnewline
 	FDM & Finite Difference Method\tabularnewline
 	FEM & Finite Element Method\tabularnewline
 	PA & Polynomial approximation\tabularnewline
 	PDE & Partial Differential Equation\tabularnewline
 	SG & Spectral-Galerkin\tabularnewline
  	TEC & Thermal equivalent circuit\tabularnewline
 	& \tabularnewline
 	\multicolumn{2}{l}{\emph{Greek}}\tabularnewline
 	$\nu$ & test function\tabularnewline
 	$\rho$ & density {[}kg m\textsuperscript{-3}{]}\tabularnewline
 	$\varphi$ & azimuthal coordinate \tabularnewline
 	$\phi$ & basis function\tabularnewline
 	$\Psi$ & vector of basis functions\tabularnewline
 	$\omega$ & frequency\tabularnewline
 	& \tabularnewline
 	\multicolumn{2}{l}{\emph{Subscripts}}\tabularnewline
 	b & bottom edge\tabularnewline
 	l & left edge\tabularnewline
 	r & right edge\tabularnewline
 	t & top edge\tabularnewline
	$\infty$ & ambient condition\tabularnewline
 	& \tabularnewline
 \end{supertabular}


\section*{Appendix A}
{Matlab code for the model described in this paper is available online at www.github.com/robert-richardson/Spectral-Thermal-Model-2D.}

\FloatBarrier
\section*{References}

\bibliography{elsarticle-template}

\begin{thebibliography}{10}
\expandafter\ifx\csname url\endcsname\relax
  \def\url#1{\texttt{#1}}\fi
\expandafter\ifx\csname urlprefix\endcsname\relax\def\urlprefix{URL }\fi
\expandafter\ifx\csname href\endcsname\relax
  \def\href#1#2{#2} \def\path#1{#1}\fi

\bibitem{Forgez2010a}
C.~Forgez, D.~{Vinh Do}, G.~Friedrich, M.~Morcrette, C.~Delacourt,
  \href{http://linkinghub.elsevier.com/retrieve/pii/S037877530901982X}{{Thermal
  modeling of a cylindrical LiFePO4/graphite lithium-ion battery}}, Journal of
  Power Sources 195~(9) (2010) 2961--2968.
\newblock \href {http://dx.doi.org/10.1016/j.jpowsour.2009.10.105}
  {\path{doi:10.1016/j.jpowsour.2009.10.105}}.
\newline\urlprefix\url{http://linkinghub.elsevier.com/retrieve/pii/S037877530901982X}

\bibitem{Spinner2015}
N.~S. Spinner, C.~T. Love, S.~L. Rose-Pehrsson, S.~G. Tuttle,
  \href{http://linkinghub.elsevier.com/retrieve/pii/S0013468615013602}{{Expanding
  the Operational Limits of the Single-Point Impedance Diagnostic for Internal
  Temperature Monitoring of Lithium-ion Batteries}}, Electrochimica Acta\href
  {http://dx.doi.org/10.1016/j.electacta.2015.06.003}
  {\path{doi:10.1016/j.electacta.2015.06.003}}.
\newline\urlprefix\url{http://linkinghub.elsevier.com/retrieve/pii/S0013468615013602}

\bibitem{Wang2012a}
Q.~Wang, P.~Ping, X.~Zhao, G.~Chu, J.~Sun, C.~Chen,
  \href{http://linkinghub.elsevier.com/retrieve/pii/S0378775312003989}{{Thermal
  runaway caused fire and explosion of lithium ion battery}}, Journal of Power
  Sources 208 (2012) 210--224.
\newblock \href {http://dx.doi.org/10.1016/j.jpowsour.2012.02.038}
  {\path{doi:10.1016/j.jpowsour.2012.02.038}}.
\newline\urlprefix\url{http://linkinghub.elsevier.com/retrieve/pii/S0378775312003989}

\bibitem{Li2013a}
X.~Li, F.~He, L.~Ma,
  \href{http://linkinghub.elsevier.com/retrieve/pii/S037877531300671X}{{Thermal
  management of cylindrical batteries investigated using wind tunnel testing
  and computational fluid dynamics simulation}}, Journal of Power Sources 238
  (2013) 395--402.
\newblock \href {http://dx.doi.org/10.1016/j.jpowsour.2013.04.073}
  {\path{doi:10.1016/j.jpowsour.2013.04.073}}.
\newline\urlprefix\url{http://linkinghub.elsevier.com/retrieve/pii/S037877531300671X}

\bibitem{Lin2011}
X.~Lin, H.~E. Perez, J.~B. Siegel, A.~G. Stefanopoulou, Y.~Ding, M.~P.
  Castanier, {Parameterization and Observability Analysis of Scalable Battery
  Clusters for Onboard Thermal Management}, in: Les Rencontres Scientifiques
  d’IFP Energies nouvelles, no. December, 2011.

\bibitem{Lin2014}
X.~Lin, H.~E. Perez, S.~Mohan, J.~B. Siegel, A.~G. Stefanopoulou, Y.~Ding,
  M.~P. Castanier,
  \href{http://linkinghub.elsevier.com/retrieve/pii/S0378775314001244}{{A
  lumped-parameter electro-thermal model for cylindrical batteries}}, Journal
  of Power Sources 257 (2014) 1--11.
\newblock \href {http://dx.doi.org/10.1016/j.jpowsour.2014.01.097}
  {\path{doi:10.1016/j.jpowsour.2014.01.097}}.
\newline\urlprefix\url{http://linkinghub.elsevier.com/retrieve/pii/S0378775314001244}

\bibitem{Mahamud2011a}
R.~Mahamud, {Advanced battery thermal management for electrical-drive vehicles
  using reciprocating cooling flow and spatial-resolution, lumped-capacitance
  thermal model}, Ph.D. thesis, University of Nevada (2011).

\bibitem{Damay2013}
N.~Damay, C.~Forgez, M.-P. Bichat, G.~Friedrich, A.~Ospina,
  \href{http://ieeexplore.ieee.org/lpdocs/epic03/wrapper.htm?arnumber=6699893}{{Thermal
  modeling and experimental validation of a large prismatic Li-ion battery}},
  IECON 2013 - 39th Annual Conference of the IEEE Industrial Electronics
  Society (2013) 4694--4699\href {http://dx.doi.org/10.1109/IECON.2013.6699893}
  {\path{doi:10.1109/IECON.2013.6699893}}.
\newline\urlprefix\url{http://ieeexplore.ieee.org/lpdocs/epic03/wrapper.htm?arnumber=6699893}

\bibitem{Dai2015}
H.~Dai, L.~Zhu, J.~Zhu, X.~Wei, Z.~Sun,
  \href{http://linkinghub.elsevier.com/retrieve/pii/S0378775315009891}{{Adaptive
  Kalman filtering based internal temperature estimation with an equivalent
  electrical network thermal model for hard-cased batteries}}, Journal of Power
  Sources 293 (2015) 351--365.
\newblock \href {http://dx.doi.org/10.1016/j.jpowsour.2015.05.087}
  {\path{doi:10.1016/j.jpowsour.2015.05.087}}.
\newline\urlprefix\url{http://linkinghub.elsevier.com/retrieve/pii/S0378775315009891}

\bibitem{He2015}
F.~He, L.~Ma,
  \href{http://linkinghub.elsevier.com/retrieve/pii/S0017931014010783}{{Thermal
  management of batteries employing active temperature control and
  reciprocating cooling flow}}, International Journal of Heat and Mass Transfer
  83 (2015) 164--172.
\newblock \href {http://dx.doi.org/10.1016/j.ijheatmasstransfer.2014.11.079}
  {\path{doi:10.1016/j.ijheatmasstransfer.2014.11.079}}.
\newline\urlprefix\url{http://linkinghub.elsevier.com/retrieve/pii/S0017931014010783}

\bibitem{mellor1991lumped}
P.~Mellor, D.~Roberts, D.~Turner, Lumped parameter thermal model for electrical
  machines of tefc design, in: IEE Proceedings B (Electric Power Applications),
  Vol. 138, IET, 1991, pp. 205--218.

\bibitem{Wrobel2010}
R.~Wrobel, P.~H. Mellor,
  \href{http://ieeexplore.ieee.org/lpdocs/epic03/wrapper.htm?arnumber=5512838}{{A
  General Cuboidal Element for Three-Dimensional Thermal Modelling}}, IEEE
  Transactions on Magnetics 46~(8) (2010) 3197--3200.
\newblock \href {http://dx.doi.org/10.1109/TMAG.2010.2043928}
  {\path{doi:10.1109/TMAG.2010.2043928}}.
\newline\urlprefix\url{http://ieeexplore.ieee.org/lpdocs/epic03/wrapper.htm?arnumber=5512838}

\bibitem{simpson2014general}
N.~Simpson, R.~Wrobel, P.~H. Mellor, A general arc-segment element for
  three-dimensional thermal modeling, Magnetics, IEEE Transactions on 50~(2)
  (2014) 265--268.

\bibitem{qi2014methodical}
F.~Qi, A.~Stippich, M.~Guettler, M.~Neubert, R.~W. De~Doncker, Methodical
  considerations for setting up space-resolved lumped-parameter thermal models
  for electrical machines, in: Electrical Machines and Systems (ICEMS), 2014
  17th International Conference on, IEEE, 2014, pp. 651--657.

\bibitem{simpson2014accurate}
N.~Simpson, R.~Wrobel, P.~H. Mellor, An accurate mesh-based equivalent circuit
  approach to thermal modeling, Magnetics, IEEE Transactions on 50~(2) (2014)
  269--272.

\bibitem{Kim2007}
G.~H. Kim, A.~Pesaran, R.~Spotnitz, {A three-dimensional thermal abuse model
  for lithium-ion cells}, Journal of Power Sources 170~(2) (2007) 476--489.
\newblock \href {http://dx.doi.org/10.1016/j.jpowsour.2007.04.018}
  {\path{doi:10.1016/j.jpowsour.2007.04.018}}.

\bibitem{Pesaran2002d}
A.~a. Pesaran,
  \href{http://linkinghub.elsevier.com/retrieve/pii/S0378775302002008}{{Battery
  thermal models for hybrid vehicle simulations}}, Journal of Power Sources
  110~(2) (2002) 377--382.
\newblock \href {http://dx.doi.org/10.1016/S0378-7753(02)00200-8}
  {\path{doi:10.1016/S0378-7753(02)00200-8}}.
\newline\urlprefix\url{http://linkinghub.elsevier.com/retrieve/pii/S0378775302002008}

\bibitem{Gu2000a}
W.~B. Gu, C.~Y. Wang, {Thermal-Electrochemical Modeling of Battery Systems},
  Journal of The Electrochemical Society 147~(8) (2000) 2910.
\newblock \href {http://dx.doi.org/10.1149/1.1393625}
  {\path{doi:10.1149/1.1393625}}.

\bibitem{Srinivasan2003}
V.~Srinivasan, C.~Y. Wang, {Analysis of Electrochemical and Thermal Behavior of
  Li-Ion Cells}, Journal of The Electrochemical Society 150~(1) (2003) A98.
\newblock \href {http://dx.doi.org/10.1149/1.1526512}
  {\path{doi:10.1149/1.1526512}}.

\bibitem{Shah2014a}
K.~Shah, S.~Drake, D.~Wetz, J.~Ostanek, S.~Miller, J.~Heinzel, a.~Jain,
  \href{http://linkinghub.elsevier.com/retrieve/pii/S0378775314011720}{{An
  experimentally validated transient thermal model for cylindrical Li-ion
  cells}}, Journal of Power Sources 271 (2014) 262--268.
\newblock \href {http://dx.doi.org/10.1016/j.jpowsour.2014.07.118}
  {\path{doi:10.1016/j.jpowsour.2014.07.118}}.
\newline\urlprefix\url{http://linkinghub.elsevier.com/retrieve/pii/S0378775314011720}

\bibitem{Shah2015modeling}
K.~Shah, A.~Jain, Modeling of steady-state and transient thermal performance of
  a li-ion cell with an axial fluidic channel for cooling, International
  Journal of Energy Research 39~(4) (2015) 573--584.

\bibitem{Muratori2010a}
M.~Muratori, N.~Ma, M.~Canova, Y.~Guezennec, {A Model Order Reduction Method
  for the Temperature Estimation in a Cylindrical Li-Ion Battery Cell}, ASME
  2010 Dynamic Systems and Control Conference, Volume 1 (2010) 633--640\href
  {http://dx.doi.org/10.1115/DSCC2010-4200} {\path{doi:10.1115/DSCC2010-4200}}.

\bibitem{Muratori2010b}
M.~Muratori, M.~Canova, Y.~Guezennec, G.~Rizzoni, E.~Politecnico,
  V.~Lambruschini, {A Reduced-Order Model for the Thermal Dynamics of Li-Ion
  Battery Cells}, in: 6th IFAC Symposium Advances in Automotive Control,
  Munich, Germany, 2010, pp. 10--15.

\bibitem{Kim2013}
Y.~Kim, J.~B. Siegel, A.~G. Stefanopoulou,
  \href{http://ieeexplore.ieee.org/xpls/abs\_all.jsp?arnumber=6579917}{{A
  computationally efficient thermal model of cylindrical battery cells for the
  estimation of radially distributed temperatures}}, in: American Control
  Conference (ACC), 2013, Washington, DC,, 2013, pp. 698--703.
\newline\urlprefix\url{http://ieeexplore.ieee.org/xpls/abs\_all.jsp?arnumber=6579917}

\bibitem{Kim2014b}
Y.~Kim, S.~Mohan, S.~Member, J.~B. Siegel, A.~G. Stefanopoulou, Y.~Ding, {The
  Estimation of Temperature Distribution in Cylindrical Battery Cells Under
  Unknown Cooling Conditions}, IEEE Transactions on Control System Technology
  (2014) 1--10.

\bibitem{Richardson2014}
R.~R. Richardson, P.~T. Ireland, D.~A. Howey,
  \href{http://linkinghub.elsevier.com/retrieve/pii/S0378775314006302}{{Battery
  internal temperature estimation by combined impedance and surface temperature
  measurement}}, Journal of Power Sources 265 (2014) 254--261.
\newblock \href {http://dx.doi.org/10.1016/j.jpowsour.2014.04.129}
  {\path{doi:10.1016/j.jpowsour.2014.04.129}}.
\newline\urlprefix\url{http://linkinghub.elsevier.com/retrieve/pii/S0378775314006302}

\bibitem{Richardson2015a}
R.~R. Richardson, D.~A. Howey,
  \href{http://ieeexplore.ieee.org/xpls/abs_all.jsp?arnumber=7097077&tag=1}{{Sensorless
  Battery Internal Temperature Estimation using a Kalman Filter with Impedance
  Measurement}}, IEEE Transactions on Sustainable Energy 6~(4).
\newline\urlprefix\url{http://ieeexplore.ieee.org/xpls/abs_all.jsp?arnumber=7097077&tag=1}

\bibitem{hahn2012heat}
D.~W. Hahn, M.~N. Ozisik, Heat conduction, John Wiley \& Sons, 2012.

\bibitem{Roscher2015}
V.~Roscher, M.~Schneider, P.~Durdaut, N.~Sassano, S.~Pereguda, E.~Mense, K.-R.
  Riemschneider, Synchronisation using wireless trigger-broadcast for impedance
  spectroscopy of battery cells, in: Sensors Applications Symposium (SAS), 2015
  IEEE, IEEE, 2015, pp. 1--6.

\bibitem{Shen2011}
J.~Shen, T.~Tang, L.-L. Wang, {Spectral Methods: Algorithms, Analysis and
  Applications}, Springer, 2011.

\bibitem{trefethen2000spectral}
L.~N. Trefethen, Spectral methods in MATLAB, Vol.~10, Siam, 2000.

\bibitem{Subramanian2005}
V.~R. Subramanian, V.~D. Diwakar, D.~Tapriyal,
  \href{http://jes.ecsdl.org/cgi/doi/10.1149/1.2032427}{{Efficient Macro-Micro
  Scale Coupled Modeling of Batteries}}, Journal of The Electrochemical Society
  152~(10) (2005) A2002.
\newblock \href {http://dx.doi.org/10.1149/1.2032427}
  {\path{doi:10.1149/1.2032427}}.
\newline\urlprefix\url{http://jes.ecsdl.org/cgi/doi/10.1149/1.2032427}

\bibitem{Northrop2011a}
P.~W.~C. Northrop, V.~Ramadesigan, S.~De, V.~R. Subramanian,
  \href{http://jes.ecsdl.org/cgi/doi/10.1149/2.058112jes}{{Coordinate
  Transformation, Orthogonal Collocation, Model Reformulation and Simulation of
  Electrochemical-Thermal Behavior of Lithium-Ion Battery Stacks}}, Journal of
  The Electrochemical Society 158~(12) (2011) A1461.
\newblock \href {http://dx.doi.org/10.1149/2.058112jes}
  {\path{doi:10.1149/2.058112jes}}.
\newline\urlprefix\url{http://jes.ecsdl.org/cgi/doi/10.1149/2.058112jes}

\bibitem{Cai2012a}
L.~Cai, R.~E. White,
  \href{http://linkinghub.elsevier.com/retrieve/pii/S0378775312010439}{{Lithium
  ion cell modeling using orthogonal collocation on finite elements}}, Journal
  of Power Sources 217 (2012) 248--255.
\newblock \href {http://dx.doi.org/10.1016/j.jpowsour.2012.06.043}
  {\path{doi:10.1016/j.jpowsour.2012.06.043}}.
\newline\urlprefix\url{http://linkinghub.elsevier.com/retrieve/pii/S0378775312010439}

\bibitem{bizeray2015lithium}
A.~Bizeray, S.~Zhao, S.~Duncan, D.~Howey, Lithium-ion battery
  thermal-electrochemical model-based state estimation using orthogonal
  collocation and a modified extended kalman filter, Journal of Power Sources
  296 (2015) 400--412.

\bibitem{Northrop2015}
P.~W.~C. Northrop, M.~Pathak, D.~Rife, S.~De, S.~Santhanagopalan, V.~R.
  Subramanian,
  \href{http://jes.ecsdl.org/cgi/doi/10.1149/2.0341506jes}{{Efficient
  Simulation and Model Reformulation of Two-Dimensional Electrochemical Thermal
  Behavior of Lithium-Ion Batteries}}, Journal of the Electrochemical Society
  162~(6) (2015) A940--A951.
\newblock \href {http://dx.doi.org/10.1149/2.0341506jes}
  {\path{doi:10.1149/2.0341506jes}}.
\newline\urlprefix\url{http://jes.ecsdl.org/cgi/doi/10.1149/2.0341506jes}

\bibitem{Doha2012a}
E.~H. Doha, a.~H. Bhrawy,
  \href{http://dx.doi.org/10.1016/j.camwa.2011.12.050}{{An efficient direct
  solver for multidimensional elliptic Robin boundary value problems using a
  Legendre spectral-Galerkin method}}, Computers and Mathematics with
  Applications 64~(4) (2012) 558--571.
\newblock \href {http://dx.doi.org/10.1016/j.camwa.2011.12.050}
  {\path{doi:10.1016/j.camwa.2011.12.050}}.
\newline\urlprefix\url{http://dx.doi.org/10.1016/j.camwa.2011.12.050}

\bibitem{fleckenstein2013thermal}
M.~Fleckenstein, S.~Fischer, O.~Bohlen, B.~B{\"a}ker, Thermal impedance
  spectroscopy-a method for the thermal characterization of high power battery
  cells, Journal of Power Sources 223 (2013) 259--267.

\bibitem{Jarrett2011a}
A.~Jarrett, I.~Y. Kim,
  \href{http://linkinghub.elsevier.com/retrieve/pii/S0378775311013279}{{Design
  optimization of electric vehicle battery cooling plates for thermal
  performance}}, Journal of Power Sources 196~(23) (2011) 10359--10368.
\newblock \href {http://dx.doi.org/10.1016/j.jpowsour.2011.06.090}
  {\path{doi:10.1016/j.jpowsour.2011.06.090}}.
\newline\urlprefix\url{http://linkinghub.elsevier.com/retrieve/pii/S0378775311013279}

\bibitem{richardson2016on}
R.~R. Richardson, S.~Zhao, D.~A. Howey, {On-board monitoring of 2-D
  spatially-resolved temperatures in cylindrical lithium-ion batteries: Part
  II. State estimation via impedance-based temperature sensing}, Arxiv Preprint
  (2016).

\end{thebibliography}

\end{document}